\documentclass[journal]{IEEEtran}

\usepackage{epsfig}
\usepackage{graphicx}
\usepackage{amsmath}
\usepackage{amsfonts}
\usepackage{amssymb}
\usepackage{mathrsfs}
\usepackage{subfigure}
\usepackage{xcolor}
\usepackage{caption}
\usepackage{float}
\usepackage{flushend}

\usepackage[pagebackref=true,breaklinks=true,colorlinks,bookmarks=false]{hyperref}

\begin{document}

\title{Neural Image Compression via Non-Local Attention Optimization and Improved Context Modeling}

\author{Tong Chen, Haojie Liu, Zhan Ma,  Qiu Shen, Xun Cao, and Yao Wang\thanks{T. Chen and H. Liu contributed equally to this work.}}
\maketitle

\begin{abstract}

This paper proposes a novel Non-Local Attention optmization and Improved Context modeling-based  image compression (NLAIC) algorithm, which is built on top of the deep nerual network (DNN)-based variational auto-encoder (VAE) structure.  Our NLAIC  1) embeds {\it non-local network operations} as non-linear transforms in the encoders and decoders for both the 
image and the latent representation probability information (known as hyperprior) to capture both local and global correlations, 2) applies {\it attention mechanism} to generate masks that are used to weigh the features, which implicitly adapt bit allocation for feature elements based on their importance, and 3) implements the improved conditional entropy modeling of latent features using {\it joint 3D convolutional neural network (CNN)-based autoregressive contexts and hyperpriors}.  
Towards the practical application, additional enhancements are also introduced to speed up processing (e.g., parallel 3D CNN-based context prediction), reduce memory consumption (e.g., sparse non-local processing) and alleviate the implementation complexity (e.g., unified model for variable rates without re-training).
The proposed model outperforms
    existing methods on Kodak and CLIC datasets with the state-of-the-art compression efficiency reported, including learned and conventional (e.g., BPG, JPEG2000, JPEG) image compression methods, for both PSNR and MS-SSIM distortion metrics.
\end{abstract}

\begin{IEEEkeywords}
	Non-local network, attention mechanism, conditional probability prediction, variable-rate model, end-to-end learning
\end{IEEEkeywords}

\section{Introduction}

Light reflected from the object surface, travels across the 3D environment and finally reaches at the sensor plane of the camera or the retina of our Human Visual System (HVS) as a projected  2D image to represent the natural scene.  Nowadays, images spread everywhere via social networking (e.g., Facebook, WeChat), professional photography sharing (e.g., flickr), online advertisement (e.g., Google Ads) and so on, mostly in standard compliant formats compressed using JPEG~\cite{wallace1991jpeg}, JPEG2000~\cite{JPEG2K}, H.264/AVC~\cite{AVC} or High Efficiency Video Coding (HEVC)~\cite{HEVC} intra-picture coding based image profile (a.k.a., Better Portable Graphics - BPG \url{https://bellard.org/bpg/}), etc. A better compression method\footnote{Since the focus of this paper is lossy compression, for simplicity, we use ``compression'' to represent ``lossy compression'' in short throughout this work, unless pointed out specifically.} is always desired to preserve the higher image quality but with less bits consumption. This would save the file storage at the Internet scale (e.g., $>$ 350 million images submitted and shared to Facebook per day), and enable the faster and more efficient image sharing/exchange with better quality of experience (QoE).

Fundamentally, image coding/compression is trying to exploit signal redundancy and represent the original pixel samples (in RGB or other color space such as YCbCr) using a compact and high-fidelity format. This is also referred to as the {\it source coding}~\cite{berger2003rate}.  Conventional transform coding (e.g., JPEG, JPEG 2000) or hybrid transform/prediction coding (e.g., intra coding of H.264/AVC and HEVC) is utilized. Here, typical transforms are Discrete Cosine Transform (DCT)~\cite{DCT}, Wavelet Transform~\cite{J2K_Wavelet}, and so on. Transforms referred here are usually with fixed basis, that are trained in advance presuming the knowledge of the source signal distribution.  On the other hand, intra prediction usually leverages the local~\cite{lainema2012intra} and global correlations~\cite{IBC} to exploit the redundancy. Since intra prediction can be expressed as the linear superimposition of casual samples, it can be treated as an alternative transform as well.
Lossy compression is then achieved via applying the quantization on transform coefficients followed by an adaptive entropy coding. 
Thus, typical image compression pipeline can be simply illustrated by ``transform'', ``quantization'' and ``entropy coding'' consecutively.

Instead of applying the handcrafted components in existing image compression methods, such as DCT, scalar quantization, etc, 
most recently emerged machine learning based image compression algorithms~\cite{balle2018variational,rippel2017real,mentzer2018conditional} leverage the autoencoder structure, which transforms raw pixels into compressible latent features via stacked convolutional neural networks (CNNs) in a nonlinear means~\cite{balle2018efficient}.  These latent features are quantized and entropy coded subsequently by further exploiting the statistical redundancy. Recent works have revealed that compression efficiency can be improved when exploring the conditional probabilities via the contexts of autoregressive spatial-channel neighbors and hyperpriors~\cite{mentzer2018conditional,li2017learning,balle2018variational,liu2018non} for the compression of features. Typically, rate-distortion optimization (RDO)~\cite{sullivan1998rate} is fulfilled by minimizing  Lagrangian cost $J$ = $R$ + $\lambda$$D$, when performing the end-to-end learning. Here, $R$ is referred to as {\it entropy rate}, and $D$ is the {\it distortion} measured by either mean squared error (MSE), multiscale structural similarity (MS-SSIM)~\cite{wang2003multiscale}, even feature or adversarial loss~\cite{simonyan2014very,huang2019extreme}.

\begin{figure*}[t]
   \centering
   \subfigure[]{\includegraphics[scale=0.55]{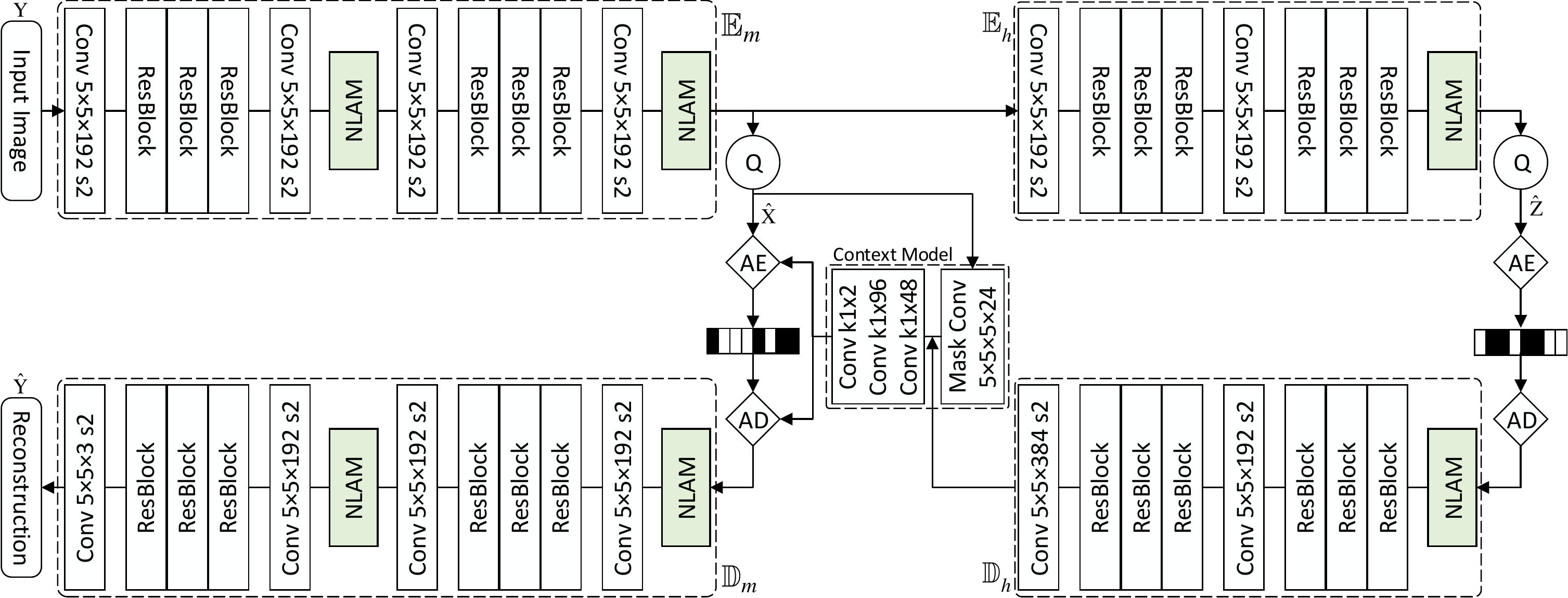}  \label{fig:framework}}
   \subfigure[]{\includegraphics[scale=0.32]{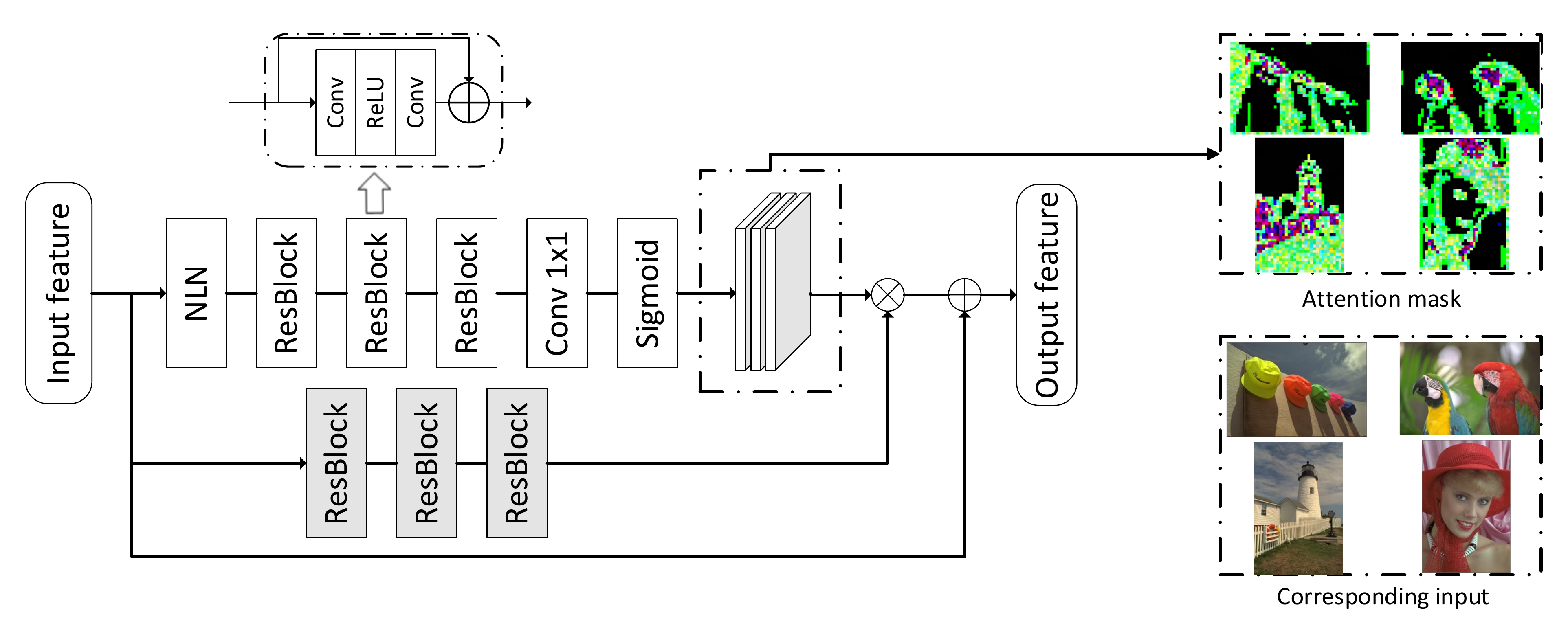}\label{sfig:non_local_attention}}
   \subfigure[]{\includegraphics[scale=0.32]{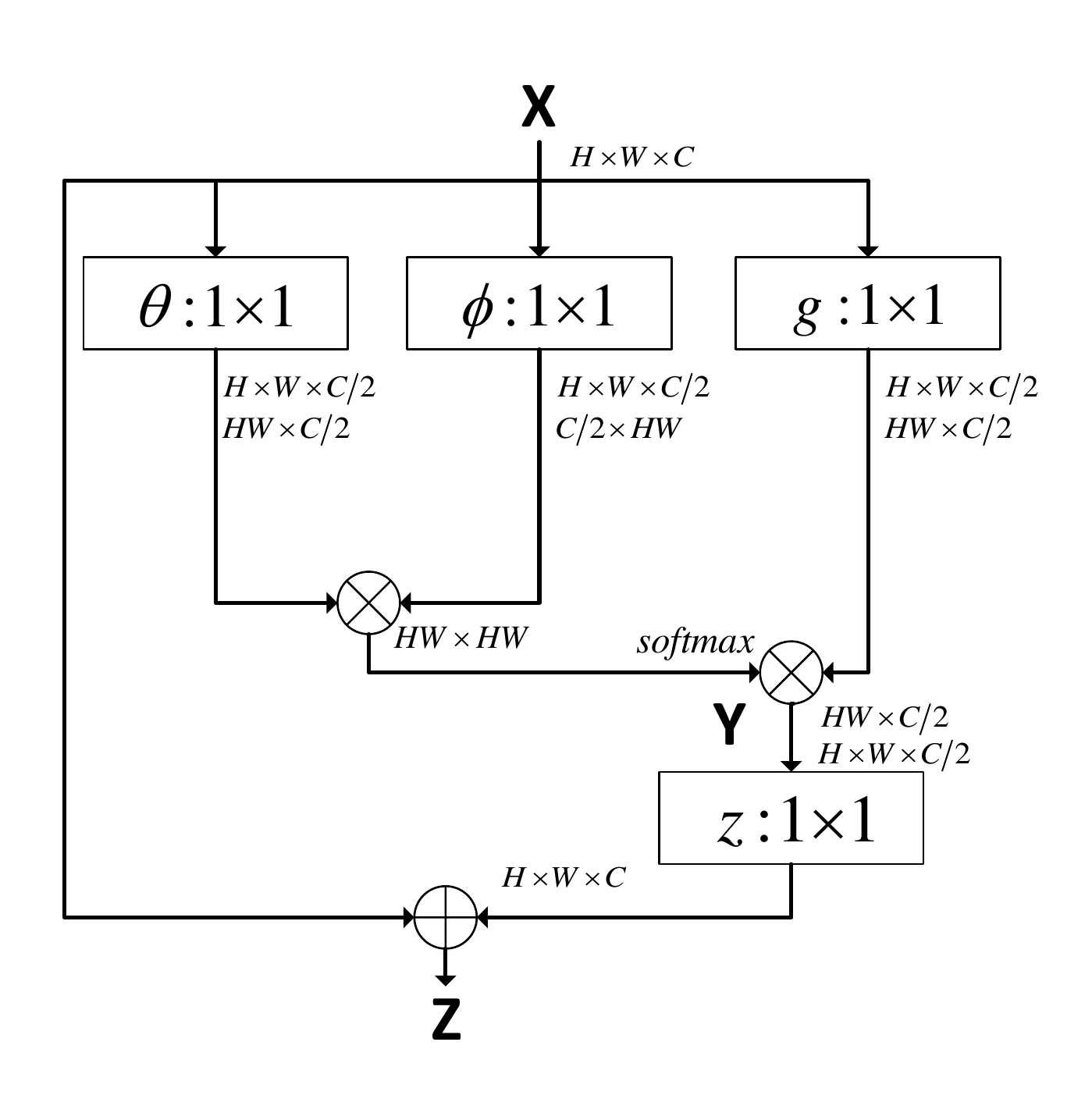} \label{sfig:non_local_fig}}
   \caption{{\bf Non-Local Attention optimization and Improved Context modeling-based image compression - NLAIC.} (a) {\it NLAIC}:  a variational autoencoder  with embedded non-local attention optimization in the main and hyperprior encoders and decoders (e.g., $\mathbb{E}_m$, $\mathbb{E}_h$, $\mathbb{D}_m$, and $\mathbb{D}_h$). "Conv 5$\times$5$\times$192 s2" indicates a convolution layer using a kernel of size 5$\times$5, 192 output channels, and stride of 2 (in decoder $\mathbb{D}_m$ and $\mathbb{D}_h$, "Conv" indicates transposed convolution). NLAM represents the Non-Local Attention Modules.  ``Q'' is for quantization, ``AE'' and ``AD'' are arithmetic encoding and decoding, $\mathbb{P}$ here denotes the probability model serving for arithmetic coding, $k1$ in context model means 3d conv kernel of size 1$\times$1$\times$1; (b) {\it NLAM}: The main branch consists of three ResBlocks. The mask branch combines non-local modules with ResBlocks for attention mask generation. The details of ResBlock is shown in the dash frame. (c) {\it Non-local network (NLN)}: $H\times{W}\times{C}$ denotes the size of feature maps with height $H$, width $W$ and channel $C$. $\oplus$ is the add operation and $\otimes$ is the matrix multiplication. }
\end{figure*}

However, existing methods still present several limitations. For example, most of the operations, such as stacked convolutions, are performed locally with limited receptive field, even with pyramidal decomposition. Furthermore, latent features are mostly treated with equal importance in either spatial or channel dimension, without considering the diverse visual sensitivities to  various content at different frequency~\cite{marrVision}. Thus, attempts have been made in~\cite{li2017learning,mentzer2018conditional} to exploit importance maps on top of the quantized latent feature vectors for adaptive bit allocation, {but still only at bottleneck layer}. These methods usually signal the importance maps explicitly. 
If the importance map is not embedded explicitly, coding efficiency will be slightly affected because of the  probability estimation error reported in~\cite{mentzer2018conditional}. 


In this paper, our NLAIC introduces {\it non-local processing} blocks into the variational autoencoder (VAE) structure to capture both local and global correlations among pixels. Attention mechanism is also embedded to generate more compact representation for both latent features and hyperpriors. Simple rectified linear unit (ReLU) is applied for nonlinear activation.  Different from those existing methods in~\cite{li2017learning,mentzer2018conditional}, our non-local attention masks are applied at  different layers (not only for quantized features at the bottleneck), to mask and adapt intelligently through the end-to-end learning framework. We also improve the context modeling of the entropy engine for better latent feature compression, by using a masked 3D CNN (i.e., 5$\times$5$\times$5) based prediction to approximate more accurate conditional statistics.

Even with the coding efficiency outperforming most existing traditional image compression standards, recent learning-based methods~\cite{minnen2018joint,balle2018variational,mentzer2018conditional} are still far from the massive adoption in reality. For practical application, compression algorithms need to be carefully evaluated and justified by its coding performance, computational and space complexity (e.g., memory consumption), hardware implementation friendliness, etc.
Few researches~\cite{rippel2017real,balle2018integer} were developed in this line for practical learned image compression coder. In this paper, we propose additional enhancements to simply the proposed NLAIC, including 1) a unified network model for variable bitrates using quality scaling factors; 2) sparse non-local processing for memory reduction; and 3) parallel 3D masked CNN based context modeling  for computational throughput improvement. All of these attempts have greatly reduce the space and time complexity of proposed NLAIC, with negligible sacrifice of the coding efficiency.

Our NLAIC has outperformed all existing learned and traditional image compression methods, offering the state-of-the-art coding efficiency, in terms of the rate distortion performance for the distortion measured by both MS-SSIM and PSNR (Peak Signal-to-Noise Ratio). Experiments have been executed using common test datasets such as Kodak~\cite{kodakset} and CLIC testing samples that are widely studied in~\cite{balle2018variational,minnen2018joint,mentzer2018conditional}. When compared with the same JPEG anchors, our NLAIC shows BD-Rate gains at 64.39\%, followed by Minnen2018~\cite{minnen2018joint} at 59.84\%, BPG (YCbCr 4:4:4) HM at 59.46\%, Ball\'e2018~\cite{balle2018variational} at 56.19\%, and JPEG2000 at 38.02\%, respectively.

Additional ablation studies have been conducted to analyze different components of the proposed NLAIC, including the impacts of sparse non-local processing, parallel 3D context modeling, unified multi-rate model, non-local operations, etc, on the coding performance, and system complexity (e.g., time and space). These investigations further demonstrate the efficiency of our NLAIC for potential practical applications.

{\bf Contributions.} We highlight the novelties of this paper below:
\begin{itemize}
   \item We are the {\it first} to introduce {\it non-local operations} into compression framework to capture both local and global correlations among the pixels in the original image and latent features.
   \item We apply {\it attention mechanism} together with the non-local operations to generate implicit {\it importance masks} at various layers to guide the adaptive processing. These masks essentially help to allocate more bits to more important areas that are critical for rate-distortion efficiency.
   \item We employ a single-layer masked 3D CNN to exploit the spatial-channel correlations in the latent features, the output of which is then concatenated with  hyperpriors to estimate the conditional statistics of the latent features, enabling more efficient entropy coding.
   \item We introduce simplifications of the original NLAIC, including the sparse non-local processing, parallel masked 3D CNN-based contexts modeling, and unified model of variable rates for practical applications to reduce computational complexity, memory storage, and to improve implementation friendliness. 
\end{itemize}

The remainder of this paper proceeds as follows: Section~\ref{sec:related_work} gives a brief review of related works, while the proposed NLAIC is discussed in Section~\ref{sec:NLAIC} followed by proposed complexity-reduction options in Section~\ref{sec:prac_NLAIC}. Section~\ref{sec:experiment} evaluates the coding efficiency of proposed method in comparison to the traditional image codecs and recently emerged learning-based approaches, while Section~\ref{sec:ablation} presents the ablation studies that examine the impact of the various components of the proposed NLAIC framework on the coding efficiency and complexity. Finally, concluding remarks and future works are described in Section~\ref{sec:concluding_remarks}.

\section{Related Work} \label{sec:related_work}
In this section, we review prior works related to the non-local operations in image/video processing, attention mechanism, as well as the learned image compression algorithms.

\textbf {Non-local Operations.} Most traditional filters (such as Gaussian and mean)                process the data locally,
by using a weighted average of spatially neighboring pixels. It usually produces over-smoothed reconstructions.
Classical non-local methods for image restoration problems (e.g., low-rank modeling~\cite{gu2014weighted}, joint sparsity~\cite{mairal2009non} and non-local means~\cite{buades2005non}) have shown their superior efficiency for quality improvement by exploiting non-local correlations. Recently, non-local operations haven been extended using DNNs for video classification~\cite{wang2018non}, image restoration (e.g., denoising, artifacts removal and super-resolution)~\cite{liu2018non, zhang2018residual}, etc, yielding significant performance improvements as reported. It is also worth to point out that non-local operations have been applied in intra coding, such as the intra block copy in HEVC-based screen content compression~\cite{IBC}, by allowing the block search in current frame to exploit non-local correlations.


\textbf {Self Attention.}
Self-attention mechanism was popularized in deep learning based natural language processing (NLP)~\cite{luong2015effective,firat2016multi,vaswani2017attention}. It can be described as a mapping strategy which  queries a set of key-value pairs to an output. For example, Vaswani {\it et. al} \cite{vaswani2017attention} have proposed multi-headed attention methods which are extensively used for machine translation. For low-level vision tasks~\cite{zhang2018residual,li2017learning,mentzer2018conditional}, self-attention mechanism makes generated features with spatial adaptive activation and enables adaptive information allocation with the emphasis on more challenging areas (i.e., rich textures, saliency, etc). 

In image compression, quantized attention masks are used for adaptive bit allocation, e.g., Li {\it et. al}~\cite{li2017learning} uses three layers of local convolutions and Mentzer {\it et. al}~\cite{mentzer2018conditional} selects one of the quantized features.
Unfortunately, these  methods require the extra explicit signaling overhead. By disabling the explicit signaling, probability estimation errors are induced. 
Our model adopts  attention mechanism that is close to~\cite{li2017learning,mentzer2018conditional} but
applies multiple layers of non-local as well as convolutional operations to automatically generate attention masks from the input image. The attention masks  are applied to the temporary latent features directly to generate the final latent features to be coded.
Thus, there is no need to use extra bits to code the masks.


\textbf {Image Compression Architectures.} DNN-based image compression generally relies on well-known autoencoders.
One direction is based on recurrent neural networks (RNN) (e.g., convolutional LSTM) in~\cite{toderici2015variable,toderici2017full,johnston2018improved}. Note that these works only require a single network model for variable rates, without resorting to the re-training. An explicit spatial adaptive bit rate allocation of compression control was suggested in~\cite{johnston2018improved} to consider the content variations for better quality at the same bit rate. More advanced bit allocation schemes, such as attention driven approaches, will be discussed later.

In another avenue, recently, (non-recurrent) CNN-based approaches~\cite{balle2016end,agustsson2017soft,rippel2017real,balle2018variational,mentzer2018conditional,li2017learning,liu2018deep,minnen2018joint,liu2019practical}. have attracted more attentions in both industry and academia. 

Among them, variational autoencoder (VAE) has been proven to be an effective structure for compression initially reported in~\cite{balle2016end}. Significant advances have been developed progressively in main components including  non-linear transforms (such as convolutions plus generalized divisive normalization - GDN in~\cite{balle2016end}), differentiable quantization (such as uniform noise approximated quantization - UNAQ~\cite{balle2016end}, and soft-to-hard quantization~\cite{agustsson2017soft}),  conditional entropy probability modeling following the Bayesian generative rules (for example, via hyperpriors~\cite{balle2018variational}, and joint 2D PixelCNN-based~\cite{oord2016pixel} autoregressive contexts and hyperpriors~\cite{minnen2018joint}). Importance map-based bit rate control are studied in~\cite{li2017learning,mentzer2018conditional} to apply more weights to important area. In addition to common MSE or MS-SSIM loss used in practice, we have witnessed other loss function designs in learning to improve the image quality, such as the feature-based or adversarial loss~\cite{liu2018deep,agustsson2018generative,huang2019extreme}.

\section{NLAIC: Non-Local Attention Optimized  Image Compression} \label{sec:NLAIC}
 Referring to the VAE structure shown in Fig.~\ref{fig:framework}, we can formulate the problem as
\begin{align}
   \min J &= R_{\mathbf{\hat{X}}} + R_{\mathbf{\hat{Z}}} + \lambda\cdot\mathsf{d}\{\mathbf{Y}, \mathbf{\hat{Y}}\}, \\
    \mathbf{\hat{X}} & = \mathsf{Q}\left\{ \mathbb{W}_{eK}\odot\left(\cdots\left(\mathbb{W}_{e2}\odot\left(\mathbb{W}_{e1}\odot\mathbf{Y} \right)\right)\right)\right\},\\
    \mathbf{\hat{Y}} & = \mathbb{W}_{dK}\odot\left(\cdots\left(\mathbb{W}_{d2}\odot\left(\mathbb{W}_{d1}\odot\mathbf{\hat{X}} \right)\right)\right),\\
    R_{\mathbf{\hat{X}}} &= - \mathsf{E}\left\{\log_2p_{\mathbf{\hat{X}}|\mathbf{\hat{Z}}}(\hat{x}_n|\hat{x}_{n-1},\ldots,\hat{x}_0,\mathbf{\hat{Z})}\right\}, \label{eq:latent_feature_prob}\\
    R_{\mathbf{\hat{Z}}} &=  - \mathsf{E}\left\{\log_2p_{\mathbf{\hat{Z}}}(\hat{z}_n)\right\}. \label{eq:hyperprior_prob}
\end{align} We wish to find appropriate parameters of transforms $\mathbb{W}_{ek}$ and $\mathbb{W}_{dk}$ ($k\in[1,K]$), quantization $\mathsf{Q}\{\}$ and conditional entropy coding for better compression efficiency, in an end-to-end learning fashion. Distortion $\mathsf{d}\{\}$ between original input $\mathbf Y$ and reconstructed image $\mathbf{\hat Y}$ is measured by either MSE or negative MS-SSIM in current study. Other distortion measurements can be applied in this learning framework as well, such as feature loss~\cite{simonyan2014very}. Bitrate is estimated using the expected entropy of latent and hyperprior features. For instance, bitrate for hyperpriors $R_{\mathbf{\hat{Z}}} $ is based on the self probability distribution (e.g., \eqref{eq:hyperprior_prob}), while bit rate for latent features $R_{\mathbf{\hat{X}}} $ is derived by exploring the probability conditioned on the distribution of both autoregressive neighbors in the feature maps (e.g., causal spatial-channel neighbors) and hyperpriors (e.g., \eqref{eq:latent_feature_prob}). $\odot$ is for the convolutional operations.


Figure~\ref{fig:framework} illustrates detailed network structures and associated parameter settings of five different components in our NLAIC system. Our NLAIC is built on a variational autoencoder structure~\cite{balle2018variational}, with {\it non-local attention modules} (NLAM) as basic units in both main and hyperprior encoder-decoder pairs (i.e., ${\mathbb E}_M$, ${\mathbb D}_M$, ${\mathbb E}_h$ and ${\mathbb D}_h$). ${\mathbb E}_M$ with quantization $Q$ are used to generate quantized latent  features and ${\mathbb D}_M$ decodes the features into the reconstructed image. ${\mathbb E}_h$ and ${\mathbb D}_h$ are applied to provide side information $\mathbf{\hat z}$ about the probability distribution of quantized latent features (known as hyperpriors), to enable efficient entropy coding.
The hyperpriors as well as autoregressive spatial-channel neighbors of the latent features are then processed through the conditional context model ${\mathbb P}$ to perform conditional probability estimation for entropy coding of the quantized latent features.

The NLAM module is shown in Fig.~\ref{sfig:non_local_attention}, and explained in Sections~\ref{sec:NLM} and~\ref{sec:NLAM} below.


\subsection{Non-local Network Processing}\label{sec:NLM}

Our NLAM adopts the Non-Local Network (NLN) proposed in~\cite{wang2018non} as a basic block, as shown in Figs.~\ref{sfig:non_local_attention} and~\ref{sfig:non_local_fig}. NLN has been mainly used for image/video processing, rather compression. This NLN computes the output at $i$-th position, ${\bf Y}_i$, using a weighted average of the transformed feature values of input $\bf X$, as below:
\begin{equation}
{\bf Y}_i = \frac{1}{C({\bf X})}{\sum_{{\forall}j}}f({\bf X}_i, {\bf X}_j)g({\bf X}_j),
\label{Eq1}
\end{equation}
where $i$ is the location index of output vector $\bf Y$ and $j$ represents the index that enumerates all accessible positions of input $\bf X$. $\bf X$ and $\bf Y$ share the same size. The function $f(\cdot)$ computes the correlations between ${\bf X}_i$ and ${\bf X}_j$, and  $g(\cdot)$ derives the representation of the input at the position $j$. $C({\bf X})$ is a normalization factor to generate the final response which is set as $C({\bf X}) = \sum_{\forall j}f({\bf X}_i, {\bf X}_j)$. Note that a variety of function forms of $f(\cdot)$ have been already discussed in~\cite{wang2018non}. Thus in this work, we directly use the embedded Gaussian function for $f(\cdot)$ for simplicity, i.e.,
\begin{equation}
f({\bf X}_i, {\bf X}_j) = e^{\theta({\bf X}_i)^T\phi({\bf X}_j)}.
\label{Eq2}
\end{equation}
Here, $\theta({\bf X}_i) = W_{\theta}{{\bf X}_i}$ and $\phi({\bf X}_j) = W_{\phi}{{\bf X}_j}$, where $W_{\theta}$ and $W_{\phi}$ denote the cross-channel transform using  1$\times$1 convolution  in our framework. The weights $f({\bf X}_i,{\bf X}_j)$ are further abstracted using a {\tt softmax} operation. The operation defined in Eq.~\eqref{Eq1} can be written in matrix form~\cite{wang2018non} as:
\begin{equation}
{\bf Y} = {\tt softmax}\left({\bf X}^TW_{\theta}^TW_{\phi}{\bf X}\right)g({\bf X}). \label{eq:nl_Y}
\end{equation}
In addition, residual connection can be added for better convergence as suggested in~\cite{wang2018non}, as shown in Fig.~\ref{sfig:non_local_fig}, i.e.,
\begin{equation}
{\bf Z} = W_{z}{\bf Y}+ {\bf X}, \label{eq:nl_Z}
\end{equation}
where $W_z$ is also a linear 1$\times$1 convolution across all channels, and ${\bf Z}$ is the final output vector with augmented global and local correlation of input $\bf X$ via above NLN.
\subsection{Non-local Attention Module (NLAM)}\label{sec:NLAM}
Note that importance (attention) map has been studied in~\cite{li2017learning,mentzer2018conditional} to adaptively allocate information to quantized latent features. For instance, we can give more bits to edge area but less bits to elsewhere, resulting in better visual quality at the similar bit rate consumption. Such adaptive allocation can be implemented by using an explicit {\it mask} encapsulated in the compressed stream with additional bits. {Note that explicit mask can be implemented using implicit signaling as well, but with prediction error-induced coding efficiency degradations as reported in~\cite{mentzer2018conditional}}. 
In addition, existing mask generation methods at only bottleneck layer in~\cite{li2017learning,mentzer2018conditional} are too simple to handle areas with more complex content characteristics.


As shown in Fig.~\ref{sfig:non_local_attention}, the entire NLAM presents three branches. 
The main (or feature) branches uses conventional stacked convolutional networks (e.g., three residual blocks, a.k.a., ResBlock, in this work~\cite{he2016deep}) to generate features, and the mask branch applies the NLN, followed by another three ResBlocks, one 1$\times$1 convolution and non-linear {\tt sigmoid} activation to produce a joint {\it spatial-channel attention mask} $M$, i.e.,\begin{equation}
   {\bf M} = {\tt sigmoid}(\mathcal{F}_{\rm NLN}({\bf X})), \label{eq:attention_mask}
   \end{equation}
   where $\bf M$ denotes the attention mask and $\bf X$ is the input feature vector. ${\mathcal{F}}_{\rm NLN}(\cdot)$ represents the operations using NLN, ResBlocks and 1$\times$1 convolution in Fig.~\ref{sfig:non_local_attention}.
This attention mask $\bf M$, having its element $0<{\bf M}_k<1, {\bf M}_k\in{\mathbb{R}}$, is multiplied element-wise with corresponding pixel element in feature maps from the main branch to perform adaptive processing.
Another residual connection in Fig.~\ref{sfig:non_local_fig} is added for faster convergence~\cite{he2016deep}.

\begin{figure}[t]
   \centering
   \includegraphics[scale=0.33]{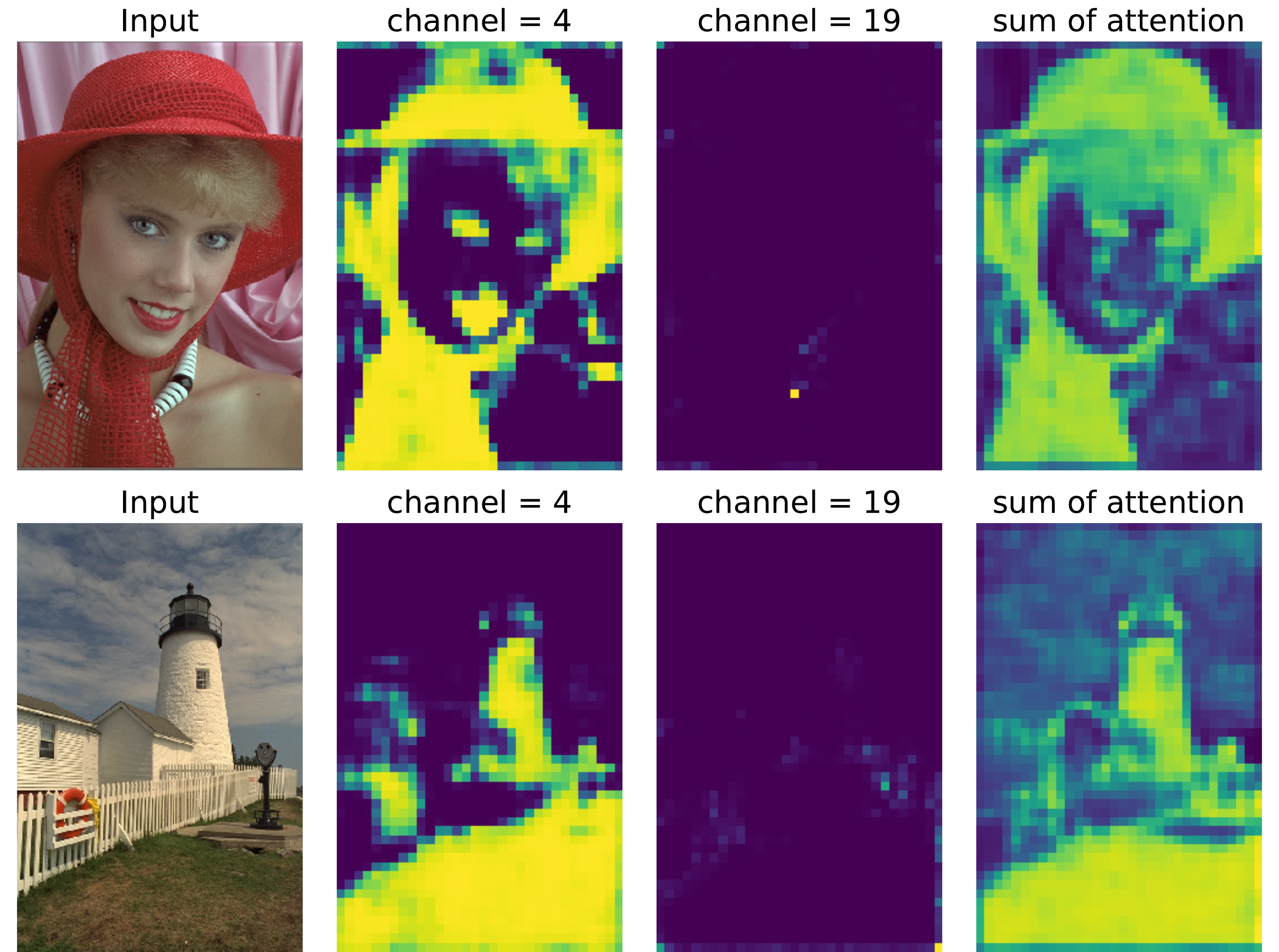}
   \caption{{\bf Visualization of attention masks generated by NLAM.} Brighter means more attention. Attention masks have the same size as latent features with several channels. Here channel 4 and 19 are picked to show the spatial and channel attention mechanism (values are in the range of $(0,1)$). Term ``sum of attention" denotes the accumulated attention maps over all channels.}
   \label{vis_attention}
\end{figure}

In comparison to the existing masking methods~\cite{li2017learning,mentzer2018conditional}, our NLAM only uses attention masks implicitly. Furthermore, multiple NLAMs (e.g., two pairs of NLAMs in the main encoder-decoder, and one pair of NLAM in the hyperprior encoder-decoder) are embedded, to massively exploit the non-local and local correlations at multi scales for accurate mask generation, rather than performing mask generation at bottleneck layer only in~\cite{li2017learning,mentzer2018conditional}. NLAMs at different layers are able to provide attention masks with multiple levels of granularity. Visualization of the attention masks generated by our NLAM can be shown in Fig.~\ref{vis_attention}. As will be reported in later experiments, NLAM embedded at various layers could offer noticeable compression performance improvement.

Since batch normalization (BN) is sensitive to the data distribution, we avoid any BN layers and only use one ReLU in our ResBlock, justified through our experimental observations. Note that in existing learned image compression methods, particularly for those with
superior performance~\cite{balle2016end,balle2018variational,minnen2018joint,liu2019gated},  GDN~\cite{balle2016end} activation has
proven its better efficiency compared with {\tt ReLU}, {\tt tanh},  {\tt sigmoid}, {\tt leakyReLU}, etc.  This may be due to the fact that
 GDN captures the global information across all feature channels  at the same pixel location. Whereas, our NLAIC shows that simple ReLU function works effectively without resorting to the GDN, owing to the fact that proposed NLAM  captures non-local correlations efficiently.



\subsection{Conditional Entropy Rate Modeling}
Previous sections present our novel NLAM scheme to transform the input pixels into more compact latent features.
This section details the entropy rate modeling that is critical for the overall rate-distortion efficiency.

\subsubsection{Context Modeling Using Hyperpriors} \label{sssec:context_z}
Similar as~\cite{balle2018variational}, a non-parametric, fully factorized density model is trained for hyperpriors $\hat{\bf z}$, which is described as:
\begin{equation}
  p_{\hat{\bf z}|\psi}(\hat{\bf z}|\psi) = {\prod_i} ( p_{{\bf z}_i|\psi^{(i)}}(\psi^{(i)})*\mathcal{U}(-\frac{1}{2},\frac{1}{2})) (\hat{\bf z}_i), \label{eq:prob_z}
 \end{equation}
where $\psi^{(i)}$ represents the parameters of each univariate distribution $p_{\hat{z}|\psi^{(i)}}$.

For quantized latent features $\hat{\bf x}$, each element $\hat{\bf x}_i$ can be modeled as a conditional Gaussian distribution as:
\begin{equation}
  p_{\hat{\bf x}|\hat{\bf z}}(\hat{\bf x}|\hat{\bf z}) = {\prod_i} (\mathcal{N}(\mu_i,{\sigma^2_i}) *\mathcal{U}(-\frac{1}{2},\frac{1}{2})) (\hat{\bf x}_i), \label{eq:prob_y_z}
 \end{equation}
where its $\mu_i$ and $\sigma_i$ are predicted using the hyperpriors $\hat{\bf z}$. We evaluate the bits of $\hat{\bf x}$ and $\hat{\bf z}$ using:
\begin{align}
  R_{\hat{\bf x}} &= - {\sum\nolimits_i} {\log_2}(p_{\hat{\bf x}_i|\hat{\bf z}_i}(\hat{\bf x}_i|\hat{\bf z}_i)), \label{eq:rate_y}\\
 R_{\hat{\bf z}} &= -{\sum\nolimits_i}  {\log_2}(p_{\hat{\bf z}_i|\psi^{(i)}}({\hat{\bf z}_i}|\psi^{(i)})). \label{eq:rate_z}
\end{align}
Usually, we take $\hat{\bf z}$ as the side information for estimating $\mu_i$ and $\sigma_i$  and $\hat{\bf z}$ only occupies a very small fraction of bits, as revealed in Section~\ref{sec:ablation} and shown in Fig.~\ref{hyper_bits_comsuming}. 

\begin{figure}[t]
   \centering
   \includegraphics[scale=0.6]{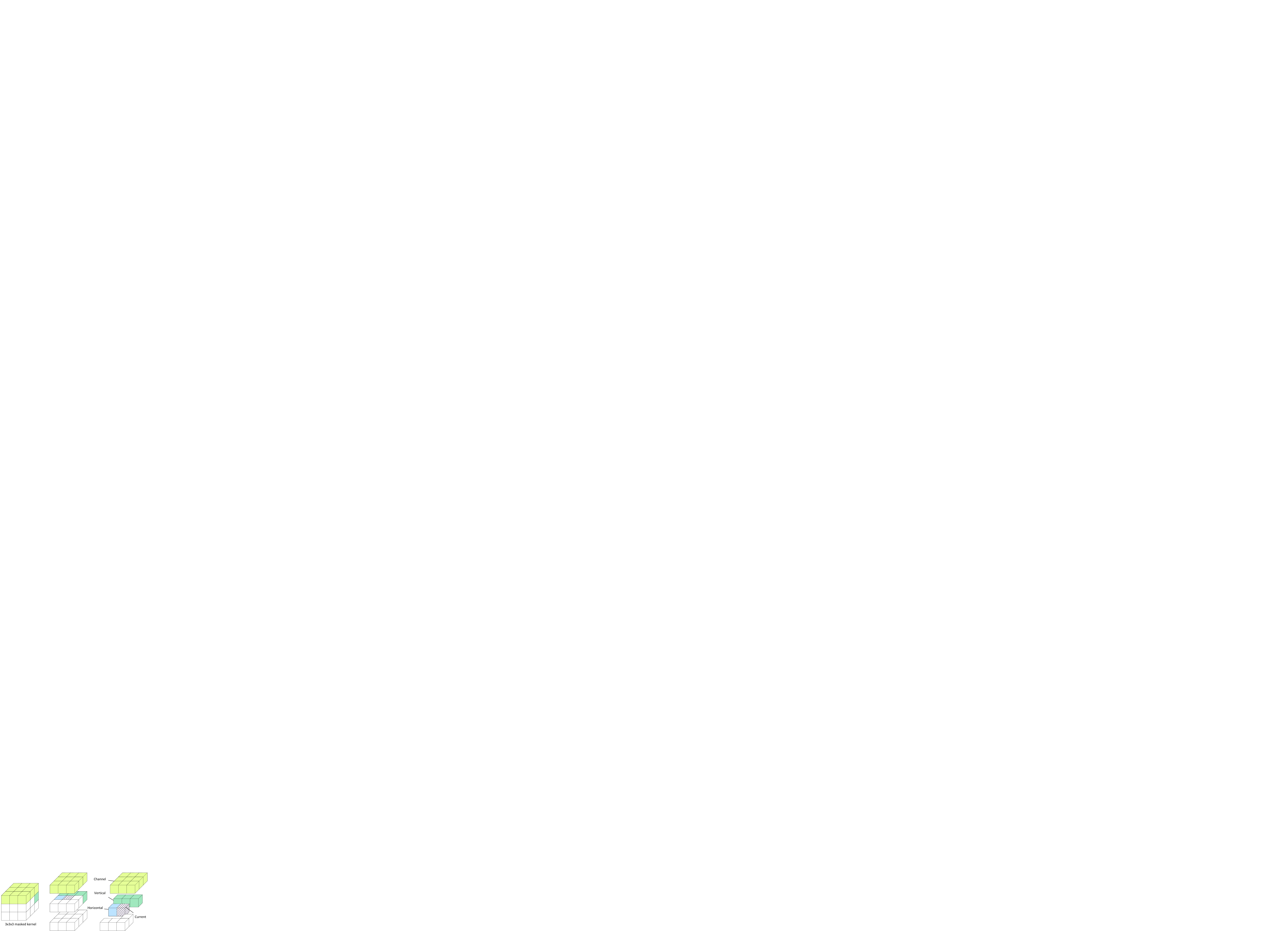}
   \caption{{\bf 3D Masked Convolution.} A 3$\times$3$\times$3 masked convolution exemplified for exploring contexts of spatial and channel neighbors jointly.  Current pixel  (in purple grid cube) is predicted by  the causal/processed pixels (in yellow for neighbors from previous channel, green for vertical neighbors,  and blue for horizontal neighbors) in a 3D space. Those unprocessed pixels (in white cube) and the current pixel are masked with zeros.}
   \label{maskconv3d}
\end{figure}

\subsubsection{Context Modeling Using Joint Autoregressive Spatial-Channel Neighbors and Hyperpriors} \label{sssec:context_joint}
Local image neighbors usually present high correlations. PixelCNNs and PixelRNNs~\cite{oord2016pixel} have been  used for effective modeling of probabilistic distribution of images using local neighbors in an autoregressive way. It is further extended for adaptive context modeling in compression framework with noticeable improvement~\cite{toderici2017full}. For example, Minnen {\it et al.}~\cite{minnen2018joint} proposed to extract autoregressive information by a 2D 5$\times$5 masked convolution at each feature channel. Such neighbor information is then combined with hyperpriors using stacked 1$\times$1 convolutions, for  probability estimation. Models in~\cite{minnen2018joint} was reported as the {\it first} learning-based method with better PSNR compared with the BPG-YUV444 at the same bit rate.

In NLAIC, we use a 5$\times$5$\times$5
3D masked convolution to exploit the spatial and cross-channel correlation jointly. This 5$\times$5$\times$5 convolutional kernel shares the same parameters for all channels, offering reduced model complexity for implementation. For simplicity, a  3$\times$3$\times$3 example is
shown in Fig.~\ref{maskconv3d}.

 Traditional 2D PixelCNNs in~\cite{minnen2018joint} need to search for a well structured channel order to exploit the conditional probability efficiently. Instead, our proposed 3D masked convolutions implicitly capture the correlations across adjacent channels.
 Compared with 2D masked CNN used in~\cite{minnen2018joint}, our 3D CNN-based approach significantly reduces the network parameters for the conditional context modeling, by enforcing the same convolutional kernels across the entire spatial-channel space.
Leveraging the additional contexts from spatial-channel neighbors via an autoregressive fashion, we can 
obtain a better conditional Gaussian distribution to model the entropy as:
\begin{align}
  p_{\hat{\bf x}}(\hat{\bf x}_i|\hat{\bf x}_1,..., &\hat{\bf x}_{i-1},\hat{\bf z})  = \nonumber\\
   & {\prod_i} (\mathcal{N}(\mu_i,{\sigma_i}^2) *\mathcal{U}(-\frac{1}{2},\frac{1}{2})) (\hat{\bf x}_i), \label{eq:context_joint}
 \end{align}
 where  $\hat{\bf x}_1, {\hat {\bf x}}_2,..., \hat{\bf x}_{i-1}$ denote the causal (and possibly reconstructed) elements prior to current $\hat{\bf x}_i$ in feature maps. $\mu_i$ and $\sigma_i$ are estimated from these causal samples and hyperpriors $\hat{\bf z}$. 



\section{Extensions of NLAIC for Complexity Reduction} \label{sec:prac_NLAIC}
This section introduces series of enhancements to reduce the model complexity of our proposed NLAIC.
For example, {\it sparse} NLAM is applied to reduce the memory consumption, while  
parallel masked 3D convolution is used to reduce the computational complexity.
Another improvement is applying a unified neural model for variable bitrates via quality mapping factors, by which the model complexity for application is greatly alleviated.

\begin{figure}[t]
	\centering
	\includegraphics[scale=0.5]{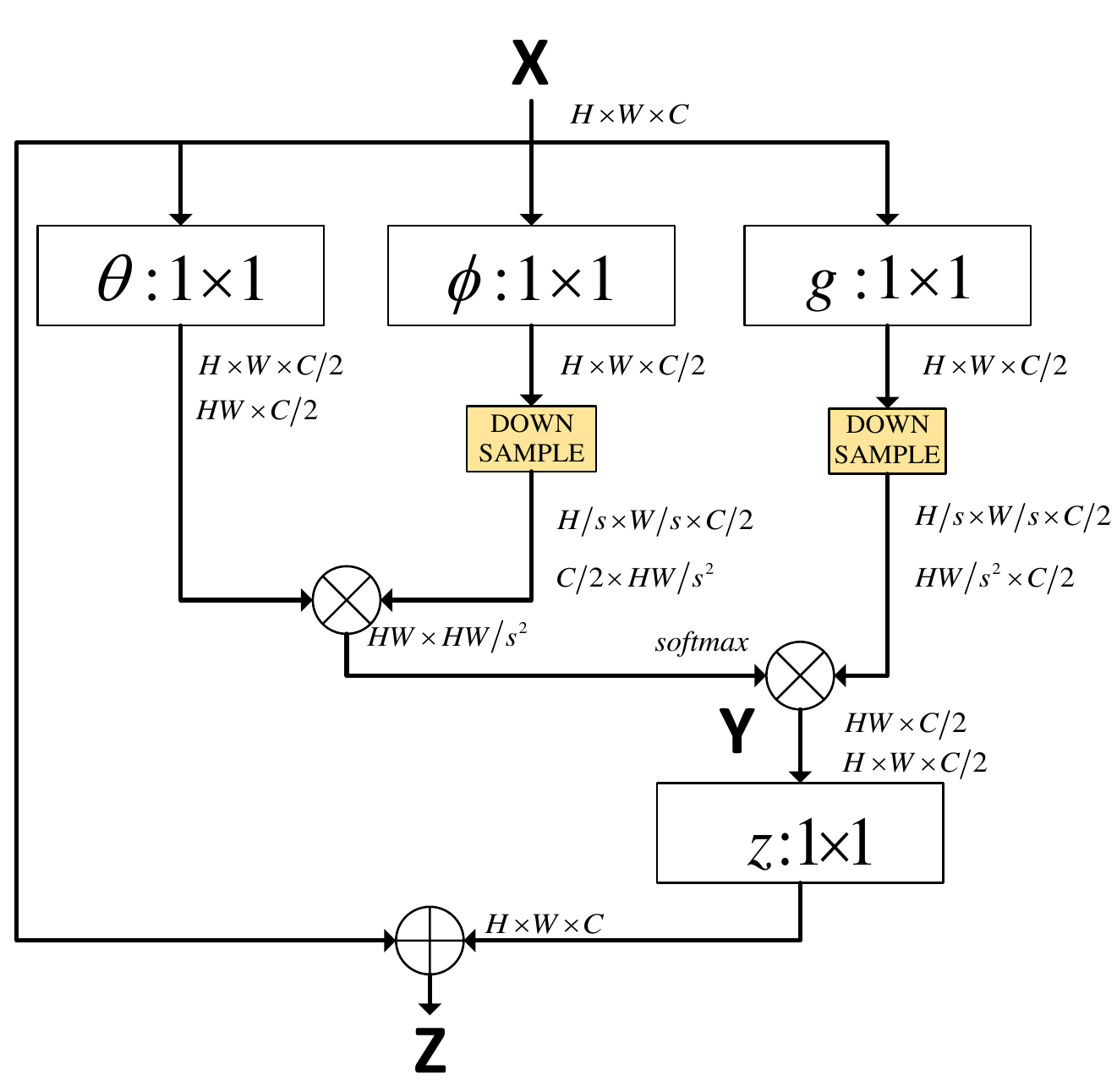}
	\caption{{\bf Sparse NLN.} Downsampling is utilized to scale down the memory consumption of the correlation matrix.}
	\label{fig:sparse_nonlocal}
\end{figure}

\subsection{Sparse NLAM} \label{ssec:sparse_NLAM}
Referring to the NLN aforementioned,  it typically requires a large amount of space/memory to host a  correlation matrix at size of $HW\times HW$ for  a na\"ive implementation. Note that $H$ and $W$ are the height and width for input feature map. 
Sparsity has been widely applied in image processing by leveraging the local pixel similarity. This motivates us to perform maxpooling-based sampling to reduce the size of the correlation matrix.
We set downsampling factor $s$ to balance between the coding efficiency and memory consumption, as shown in Fig.~\ref{fig:sparse_nonlocal}.
Our experimental studies showed that $s$ = 8 (e.g., downsampled as a factor of 64$\times$) can achieve a good trade-off. In practice, the factor $s$ can be chosen according to the memory constraints imposed by the underlying system.



\subsection{Parallel 3D Masked CNN based Context Modeling} \label{ssec:parallel_prob_modeling}
Both 2D and 3D masked convolutions~\cite{oord2016pixel} can be extended to model the conditional probability distribution for the quantized latent features pixel by pixel in Fig.~\ref{maskconv3d}.  Although the masked convolutions can leverage the neighbor pixels to predict the current pixels efficiently, it usually leads to a great computational penalty because of the strictly sequential pixel-by-pixel processing, making the compression framework far from the practical application.

Recalling the examples in Fig.~\ref{maskconv3d}, parallelism is mainly broken by using the left (horizontal) neighbor (highlighted in blue) for masked convolutions, due to a raster scan processing order. For this design, it requires $H$$\times$$W$$\times$$C$ convolutions to complete all pixel elements in feature maps, with computational complexity noted as $\mathscr{O}(H\times W\times C)$. Here, $C$ represents the number of channels of the quantized latent features. 

One simplification is to remove the dependency on left neighbors, by only using the vertical and channel neighbors. Then the convolution for each line can be performed in parallel by $W$ processors, each computing $H\times C$ convolutions. Theoretically this can reduce the processing time to $\mathscr{O}(H\times C)$.

We can further remove the vertical neighbors in Fig.~\ref{maskconv3d}, so that the convolutions for all the pixels in each feature map channel can be computed simultaneously using $H\times W$ processors, reducing the processing time to $\mathscr{O}(C)$. 
Simulations in ablation studies and in Fig.~\ref{fig:pixel} show that performance impact is negligible when we used such parallel 3D convolution-based context modeling.


\begin{figure}[t]
   \begin{center}
   \subfigure[ ]{   \includegraphics[scale=0.162]{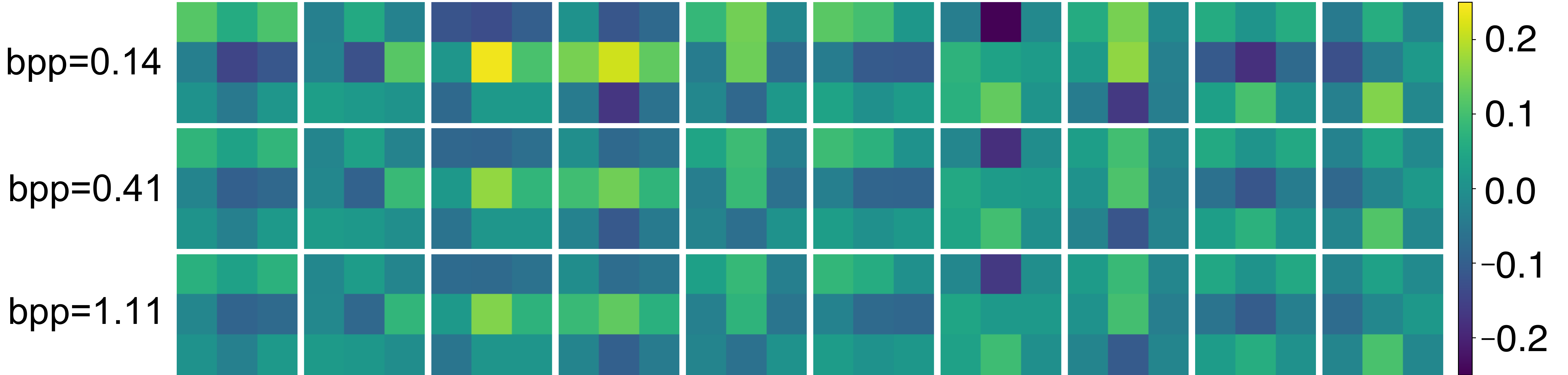}  }
   \subfigure[bpp = 0.14]{      \includegraphics[height=3.0cm]{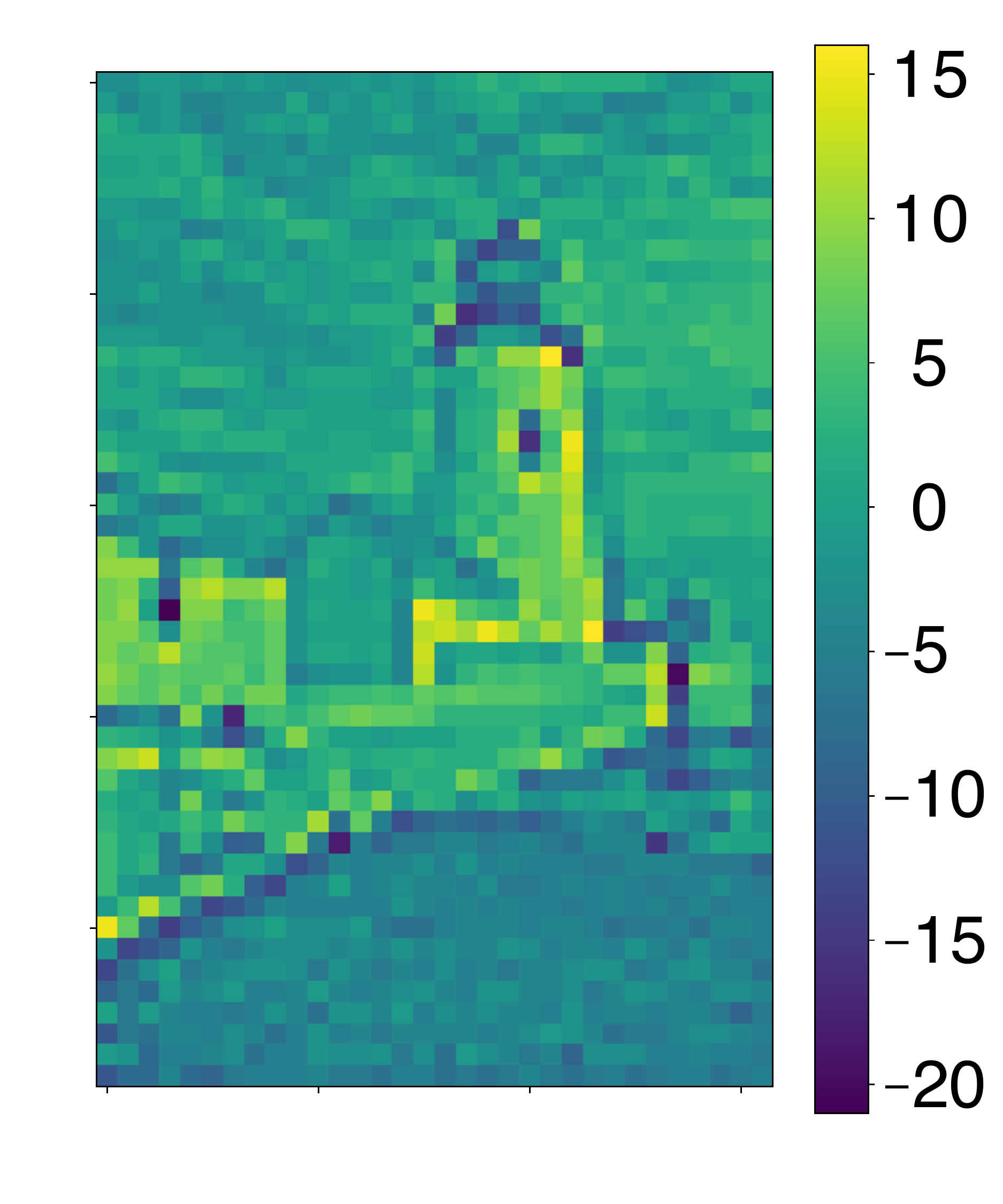}        }
   \subfigure[bpp = 0.41]{      \includegraphics[height=3.0cm]{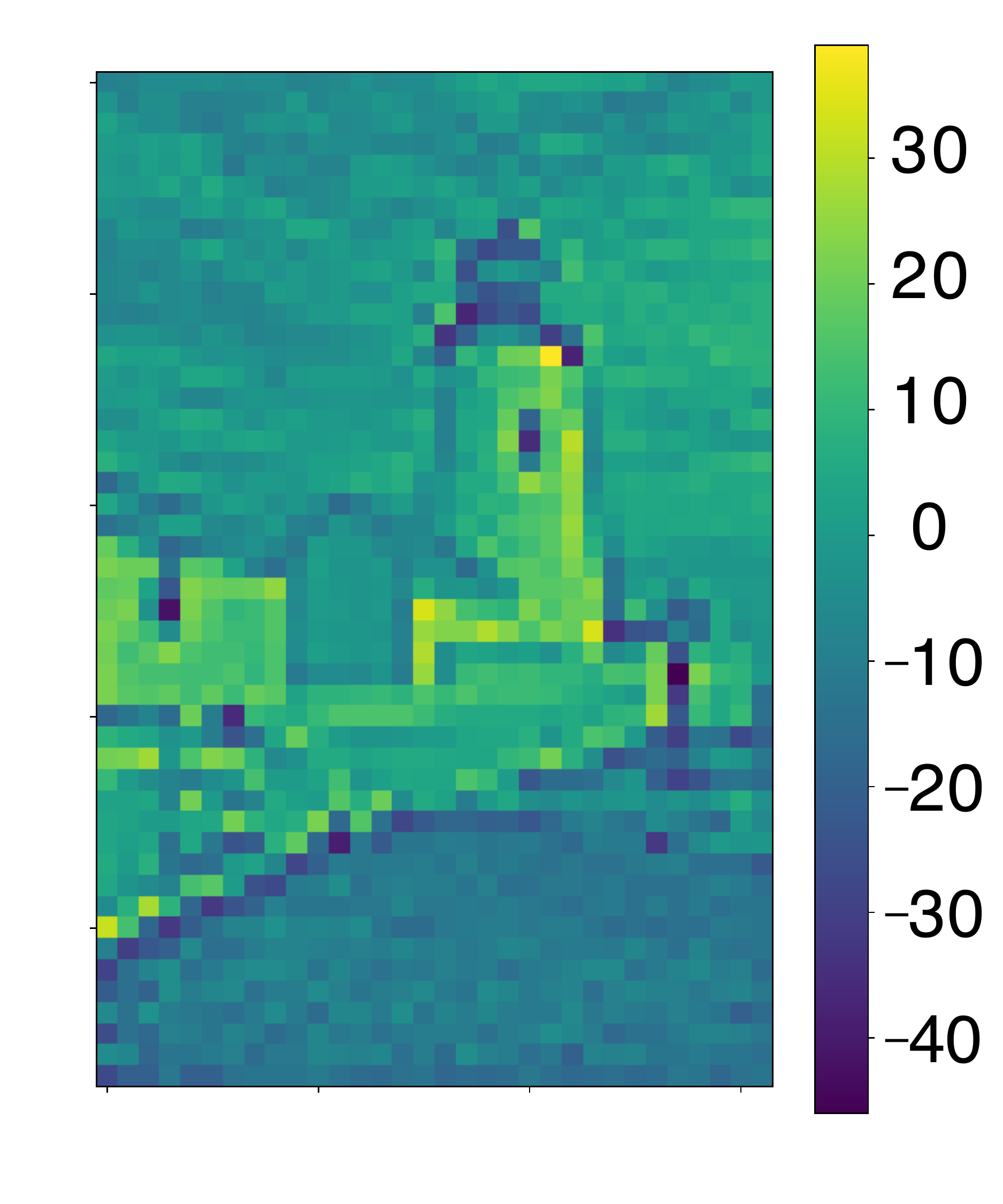}        }
   \subfigure[bpp = 1.11]{      \includegraphics[height=3.0cm]{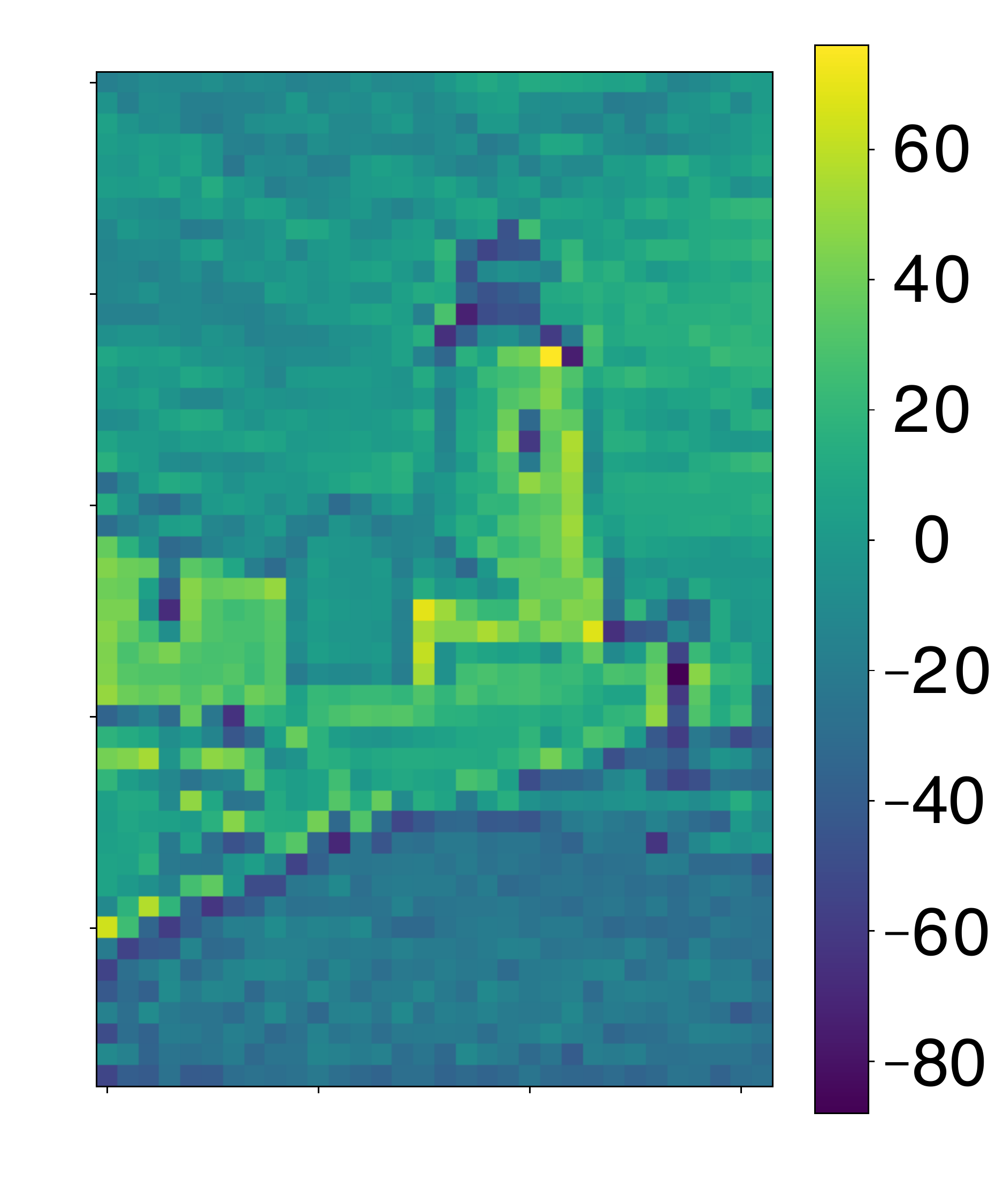}       }
   \end{center}
   \caption{{\bf Visualization at Variable Rates.} (a) Convolutional Kernels with bitrates at 0.14 bpp (first row), 0.41 bpp (second row), 1.11 bpp (last row); (b)-(d) Feature maps at corresponding bitrates.}
   \label{fig:vis_fmaps}
\end{figure}

\subsection{A Unified Model for Variable Rates}
\label{ssec:unified_model_var_rates}

Deep learning-based image coders are usually trained by minimizing a rate-distortion criteria, i.e., $R+\lambda D$, where $\lambda$ is a hyperparameter controlling the trade-off between $R$ and $D$. For most studies~\cite{balle2018variational,minnen2018joint,mentzer2018conditional}, they have trained different models for different target bitrates by varying $\lambda$. Using different models for different rates not only brings significant memory consumption to host/cache them, but also introduces additional model switching overhead when performing the bitrate adaptation in encoder optimization. 
Thus, a single or unified model covering a reasonable range of bitrate is highly desirable in practice.

A few attempts have been made in pursuit of a single/unified model for variable rates without re-training for individual bitrate, such as ~\cite{toderici2015variable,toderici2017full,johnston2018improved,zhang2018learned}, all of which compress bit-planes progressively.
These methods inherently offer the bit-depth scalability. Standing upon the scalability perspective, a layered network design~\cite{jia2019layered} is proposed to produce fine-grain bit rates in a unified framework, but it is actually a concatenation of a base layer model trained at the low bit rate and a set of refinement models trained for successively refinement. Recently, Dumas {\it et. al}~\cite{dumas2018autoencoder} have tried to learn the transform and quantization jointly in a exemplified simple network, to offer the quantization independent compression of luminance image.



As aforementioned adapting $\lambda$ yields the best coding efficiency for a specific  bitrate target. To examine how does the learnt filters and the feature maps change with the bit rate,
we first train a model at a high bitrate ($\approx$ 1bpp), and retrain additional models at a variety of bitrates (by using increasingly larger $\lambda$) based on this high-bitrate model initialization. Corresponding feature maps and convolutional kernels are presented in Fig.~\ref{fig:vis_fmaps}. It shows that both kernels and feature maps keep almost the same pattern with  the intensity of each element scaled, for models at variable rates. This suggests that we can simply apply a set of scaling factors to adapt the last feature maps to be coded which was trained at a high bitrate scenario to other bitrate ranges, without the retraining of entire model for individual rates any more. This is analogous to the {\it scaling} operations~\cite{LowComplexityTransform_AVC} used in H.264/AVC or HEVC for transformed coefficients prior to being quantized.




\begin{figure}[t]
   \begin{center}
      \includegraphics[scale=0.85]{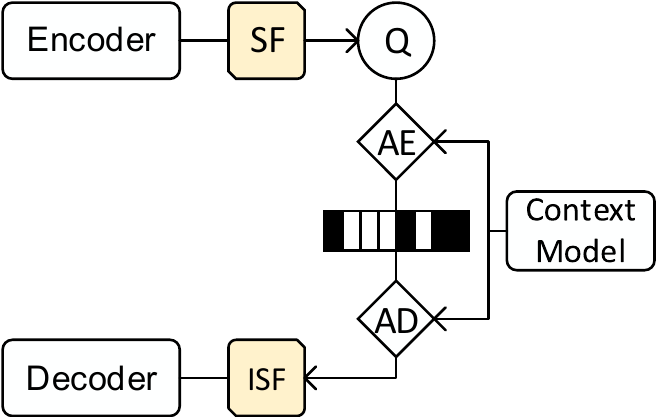}  
   \end{center}
   \caption{{\bf Quality Scaling Factor.} Feature maps are scaled using scaling factor (SF) before quantization, and inverse scaling factor (ISF) is used for entropy-decoded elements appropriately. A fixed context model is used for entropy probability estimation}
   \label{fig:qf}
\end{figure}

\begin{figure}[p]
\centering
\subfigure[]{\includegraphics[scale=0.28]{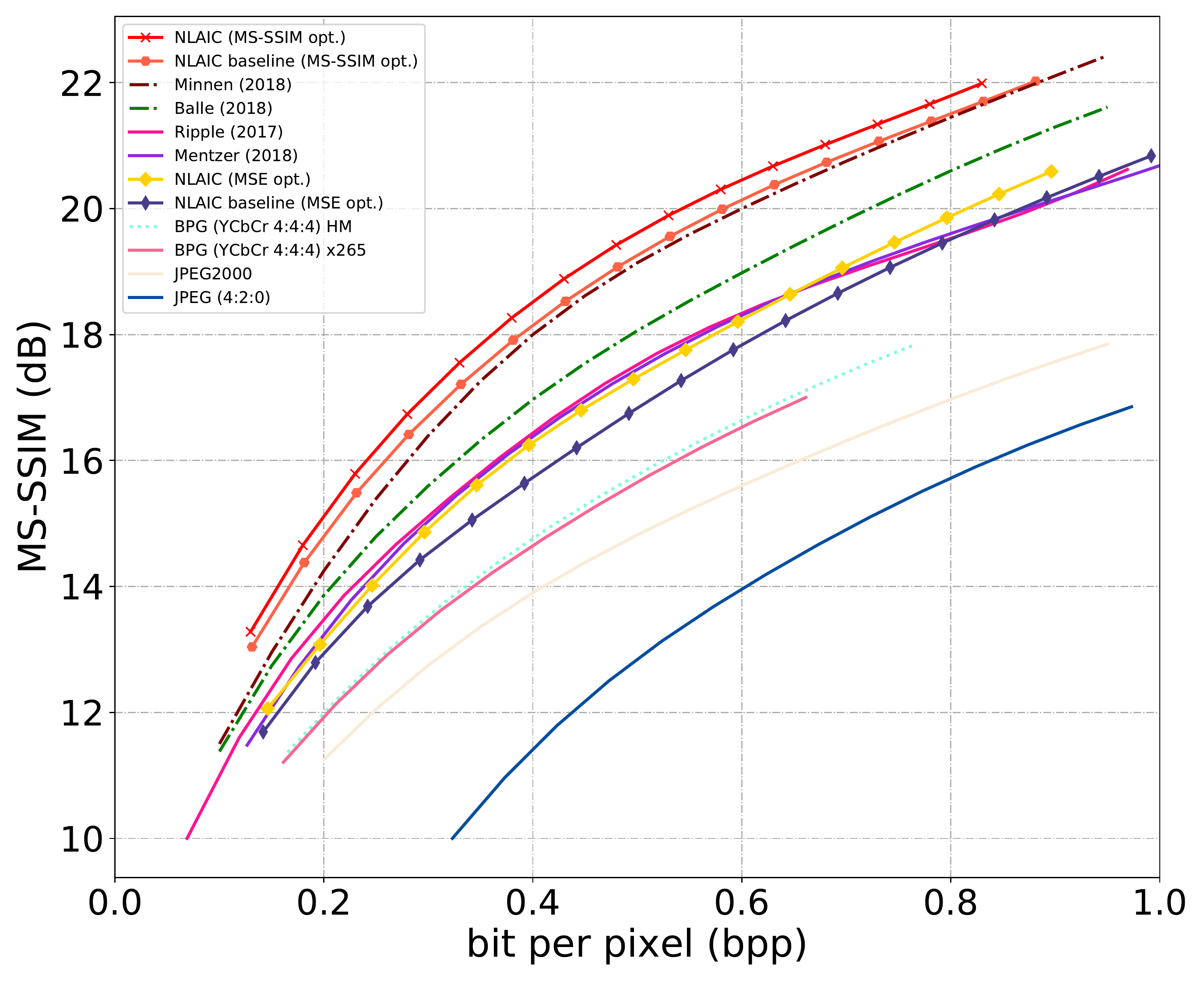} \label{sfig:ssim_perf}}\\
\subfigure[]{\includegraphics[scale=0.28]{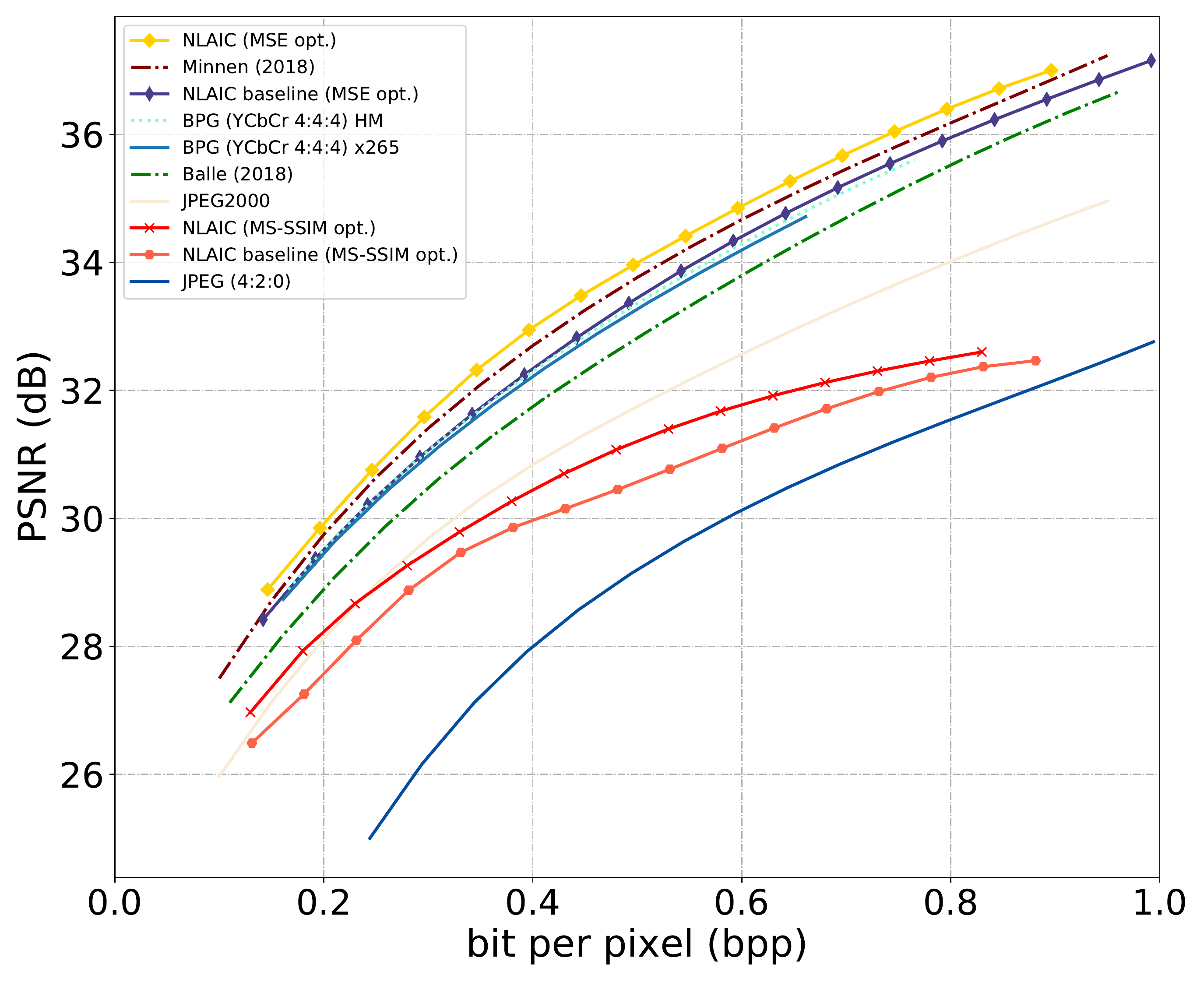} {\label{sfig:psnr_perf}}}\\
\subfigure[]{\includegraphics[scale=0.48]{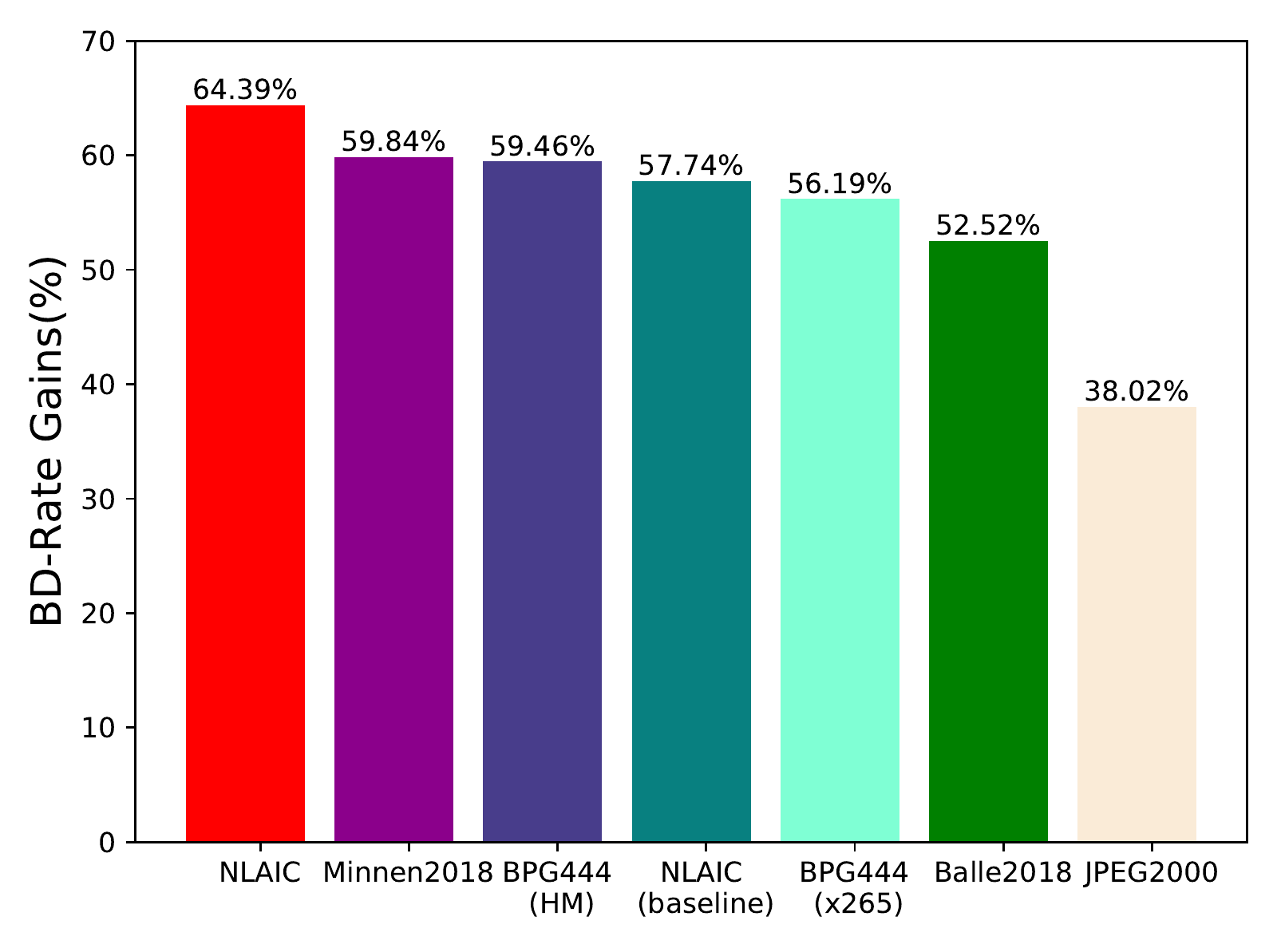}  \label{bd_rate_reduction_bar}}
\caption{{\bf Rate-distortion Efficiency}. Illustrative comparisons are given using public Kodak data. (a) distortion is measured by MS-SSIM (dB). Here we use $-10\log_{10}(1-d)$ to represent raw MS-SSIM ($d$) in dB scale. (b) PSNR is used for distortion evaluation. (c) Numerical coding gains with JPEG as anchor and distortion is measured by PSNR.}
\label{rd_curve}
\end{figure}


Thus, we propose a set of quality scaling factors (${\bf s}_f$) that will be embedded in autoencoder as shown in Fig.~\ref{fig:qf} for the bit rate adaptation. For any input image $\bf Y$, encoder $\mathbb{E}$ generates corresponding feature maps  using $ {\bf X}^0 = \mathbb{E}(\bf Y)$ at a specific bit rate, typically for a high bit-rate $R_0$. 
Scaling factors $({\bf s}_f\in \{a_0,b_0\},\{a_1,b_1\},...,\{a_n,b_n\})$ are devised to linearly scale each of all $n$ channels in input features to new bitrate. Thus, for feature maps ${\bf X}^{new}$ at new bitrate target, we have:
\begin{align}
    \hat{\bf B}_i = {\sf Q}\{ {\bf X}^{new}_i\} = {\sf Q}\{ {\sf SF}\{{\bf X^0_i}\}\} = {\sf Q}\{ a_i\cdot{\bf X}^0_i + b_i\}.
\end{align} $\hat{\bf B}_i $ is a vector of quantized elements in $i^{th}$ channel for entropy coding, and will be inversely scaled, 
\begin{align}
    \hat{\bf X}_i = {\sf ISF}\{\hat{\bf B}_i\} = \frac{\hat{\bf B}_i - b_i}{a_i},
\end{align}  prior to being fed into the decoder network $\mathbb{D}$ for reconstructed image $\hat{\bf Y} = \mathbb{D}(\hat{\bf X})$.

\begin{figure*}[t]
	\centering
	\includegraphics[scale=0.4]{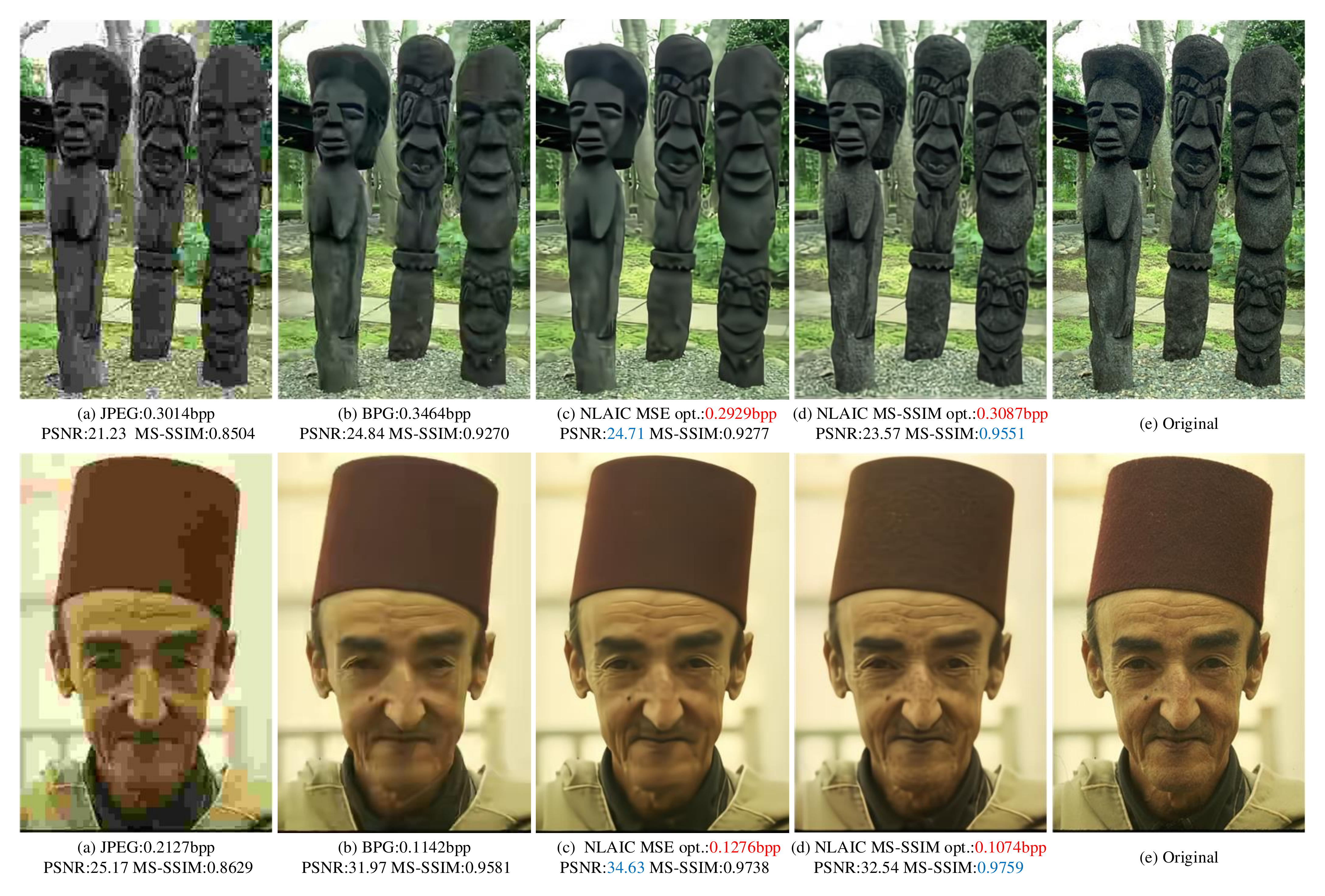}
	\caption{{\bf Subjective Evaluation.} Visual comparison among JPEG420, BPG444, NLAIC MSE opt., MS-SSIM opt. and the original image from left to right. Our method achieves the best visual quality containing more texture without blocky nor blurring artifacts. 
	}
	\label{visual_comparison}
\end{figure*}




Note that entropy context modeling $\mathbb{P}$ is actively employed in learned image compression to accurately capture the probability distribution of feature map elements for rate-distortion optimization and entropy coding. Given that elements in feature maps are scaled and biased, thus we need to adjust the quantization step in probability calculation accordingly, i.e.,

\begin{align}
      p(\hat{\bf x}) = {\prod_i} (\mathcal{N}(\mu_i,{\sigma^2_i}) *\mathcal{U}(-\frac{1}{2},\frac{1}{2})) (\hat{\bf x}_i) \\ \Rightarrow {\prod_i} (\mathcal{N}(\mu_i,{\sigma^2_i}) *\mathcal{U}(-\frac{1}{2a_i},\frac{1}{2a_i})) (\hat{\bf x}^{new}_i)
   \label{eq:estimator}
\end{align}



Though we have exemplified the quality scaling factor using an illustrative autoencoder in Fig.~\ref{fig:qf}, it can be extended to other complex network structures easily. In practice, our NLAIC implements both main and hyper encoders and decoders. However, our simulations have revealed that hyperpriors only occupy a small overhead, e.g., 2\%-8\% (larger number for smaller bpp), for an entire compressed bitstream. Thus, in the view of low-complexity and practical application, we only apply scaling factors in main codec, leaving the hyper codec fixed, which is trained at the highest bit rate.





\section{Experimental Studies}\label{sec:experiment}

This section presents comprehensive performance evaluation. Training and measurement follows the comment practices used by other compression algorithms~\cite{JPEG,JPEG2K,balle2018variational,minnen2018joint} for a fair comparison.

\subsection{Training}
 We use \textbf{COCO}~\cite{lin2014microsoft} and \textbf{CLIC}~\cite{clic}  datasets to train our NLAIC framework. We randomly crop images into  192$\times$192$\times$3 patches for subsequent learning. Well-known RDO process is applied to do end-to-end training at various bit rates via
$
L = {\lambda}{\cdot}d(\hat{\bf Y}, {\bf Y})+R_x+R_z$.
$d(\cdot)$ is a distortion measurement between reconstructed image $\hat{\bf Y}$ and the original image $\bf Y$. Both negative MS-SSIM and MSE are used in our work as distortion loss functions, which are marked as ``MS-SSIM opt.'' and ``MSE opt.'', respectively.
$R_x$ and $R_z$ represent the estimated bit rates of latent features and hyperpriors, respectively.
Note that all components of our NLAIC are trained together.
We set learning rates (LR) for ${\mathbb E}_M$, ${\mathbb D}_M$, ${\mathbb E}_h$, ${\mathbb D}_h$ and
${\mathbb P}$ at $3e{-5}$ in the beginning.
But for ${\mathbb P}$, its LR  is clipped to  $1e{-5}$  after 30 epochs.
Batch size is set to 16 and the entire model is trained on 4-GPUs in parallel.

\begin{figure*}[t]
	\centering
	\includegraphics[scale=0.3]{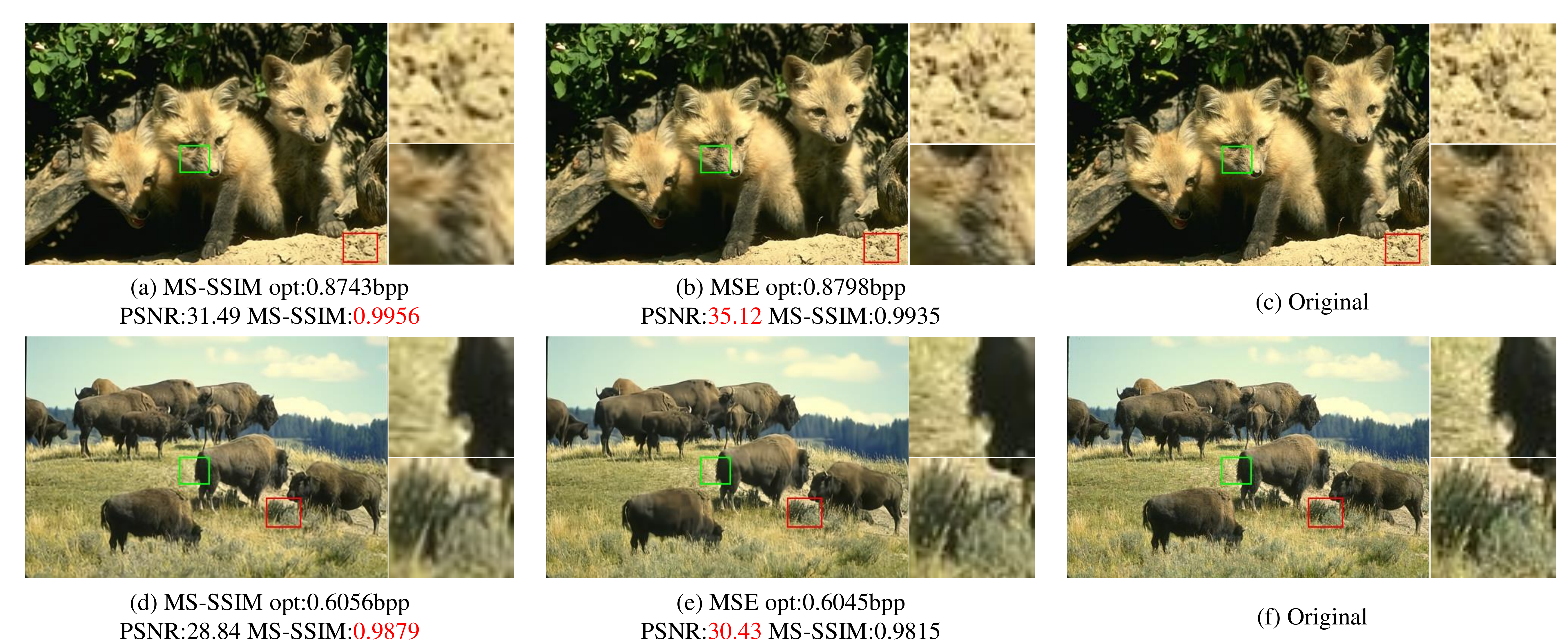}
	\caption{{\bf Impacts of Loss Functions.} Illustrative reconstruction samples of respective PSNR and MS-SSIM loss optimized compression}
	\label{psnr_ssim_comparision}
\end{figure*}


\subsection{Rate-Distortion Efficiency}
We evaluate our NLAIC models by comparing the rate-distortion performance averaged on publicly available Kodak dataset. Figure~\ref{rd_curve} shows the performance when distortion is measured by MS-SSIM and PSNR, respectively. MS-SSIM and PSNR  are widely used in image and video compression tasks.
Here, PSNR represents the pixel-level distortion while MS-SSIM describes the structural similarity. MS-SSIM is reported to offer higher correlation with human perceptual inception, especially for low bit rates~\cite{wang2003multiscale}. As we can see, our NLAIC provides the state-of-the-art performance with noticeable performance gain compared with the other existing leading methods, such as Minnen2018~\cite{minnen2018joint} and Ball{\'e}2018~\cite{balle2018variational}.

{\bf Objective Measurement.}
As shown in Fig.~\ref{sfig:ssim_perf} using MS-SSIM for both loss and final distortion measurement, and in Fig.~\ref{sfig:psnr_perf} using MSE for loss function and PSNR for final distortion, NLAIC is both ranked at the first place, offering the state-of-the-art coding efficiency. Figure~\ref{bd_rate_reduction_bar} compares the average BD-Rate reductions by various methods over the legacy JPEG encoder. Our NLAIC model shows 64.39\% and 11.97\% BD-Rate~\cite{bjontegaard2001calculation}  reduction against JPEG (4:2:0) and BPG (YCbCr 4:4:4) HM, respectively. Here BPG HM is compiled with HEVC HM reference software which has slightly better performance than x265 used in default BPG.

{\bf Subjective Evaluation.} We also evaluate our method on BSD500~\cite{bsd500} dataset, which is widely used in image restoration problems. Figure~\ref{visual_comparison} shows the results of different image codecs at the similar bit rate. Our NLAIC provides the best subjective quality with relative smaller bit rate. {In practice, some bit rate points cannot be reached for BPG and JPEG. Thus we choose the closest one to match our NLAIC bit rate.}

\subsection{Complexity Analysis}
We perform all tests on a NVIDIA P100 GPU with Pytorch toolbox. The trained model has a size of about 262MB. For an input image at a size of 512$\times$768$\times3$, The model has 291.8G FLOPs for encoder, 353.2G FLOPs for decoder and 3.46G FLOPs for context model. Forwarding encoding requires 6172MB running memory and takes about 438ms. At the decoder side, the default line-by-line decoding with context models takes most of the decoding time which can be remarkably decreased by channel-wise parallel context modeling, as discussed in Section \ref{ssec:parallel_prob_modeling}.

\section{Ablation Studies} \label{sec:ablation}
We further analyze our NLAIC in following aspects to understand the capability of our system in practice:\\

{\bf Impacts of Loss Functions.} Considering that MS-SSIM loss optimized results demonstrate much smaller PSNR at high bit rate in Fig.~\ref{sfig:ssim_perf},
we visualize decompressed images at high bit rate using models optimized for PSNR and MS-SSIM loss as shown in Fig.~\ref{psnr_ssim_comparision}. 
We find that MS-SSIM loss optimized results exhibit worse details compared with PSNR loss optimized models at high bit rate. 
This may be due to the fact that  pixel distortion becomes more significant at high bit rate, but structural similarity puts more weights at  a fairly low bit rate range. It will be interesting to explore  a better metric to cover the advantages of PSNR at high bit rate and MS-SSIM at low bit rate for an overall optimal efficiency.

 \begin{figure}[t]
   \centering
   \includegraphics[scale=0.3]{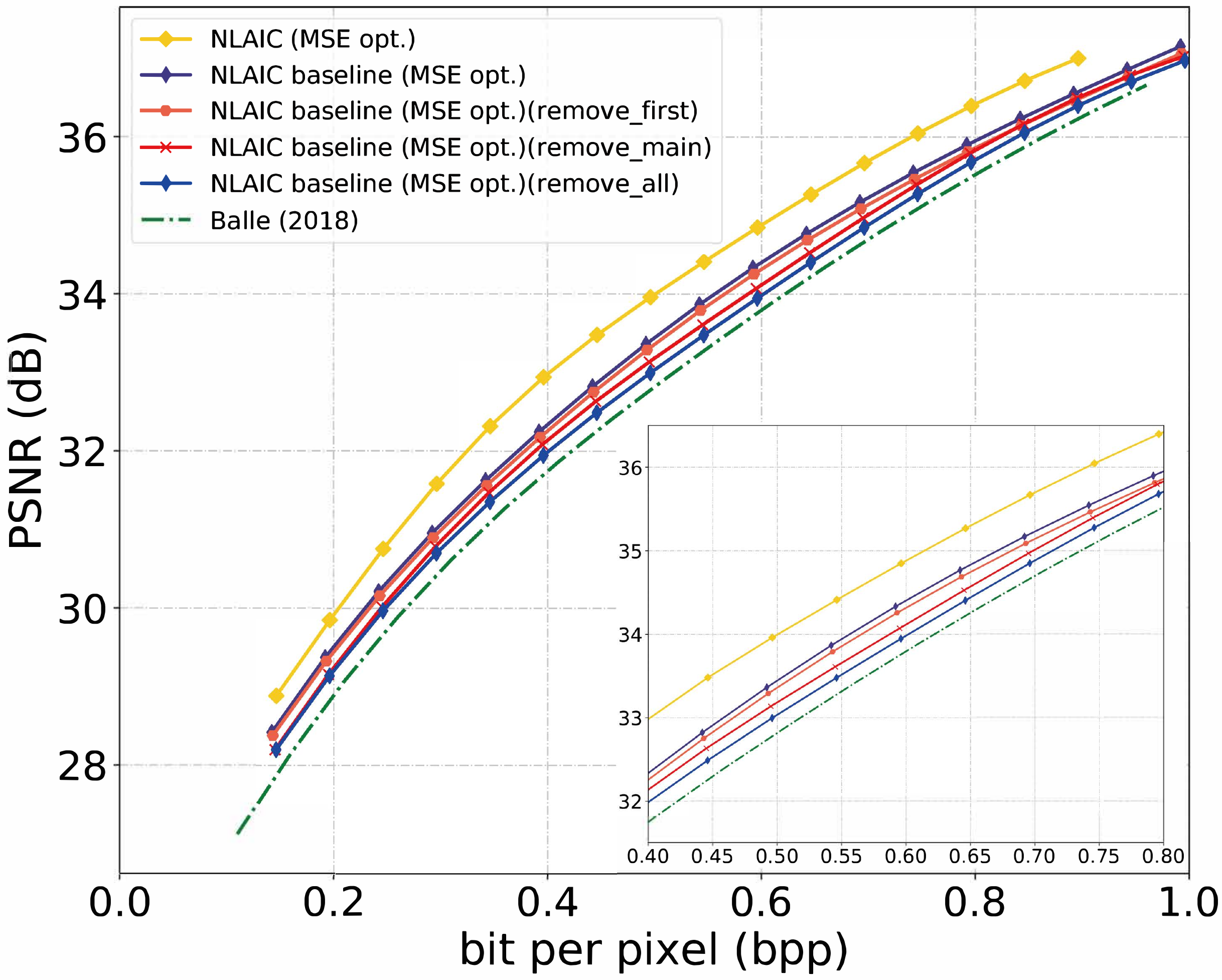}
   \caption{{\bf Impacts of NLAM.} Efficiency illustration when removing NLAM components gradually and re-training the model.}
   \label{model_variants_discuss}
\end{figure}

{\bf Impacts of Contexts.} To understand the contribution of the context modeling using spatial-channel neighbors, we offer another alternative implementation. It is referred to as ``NLAIC baseline'' that only uses the hyperpriors to estimate the means and variances of the latent features (see Eq.~\eqref{eq:prob_y_z}). In contrast, default NLAIC  uses both hyperpriors and previously coded pixels in the latent feature maps (see Eq.~\eqref{eq:context_joint}). 

Referring to Fig.~\ref{sfig:ssim_perf}, even our ``NLAIC baseline'' outperforms  the existing methods when using MS-SSIM as loss and evaluation measures. For the case that uses MSE as loss and PSNR as distortion measurement,  ``NLAIC baseline'' is slightly worse than the model in~\cite{minnen2018joint} that uses contexts from both hyperpriors and autoregressive neighbors jointly as our ``NLAIC'', but  better than the work~\cite{balle2018variational} that only uses the hyperpriors to do context modeling for fair comparison.

\begin{figure}[t]
   \centering
   \includegraphics[scale=0.18]{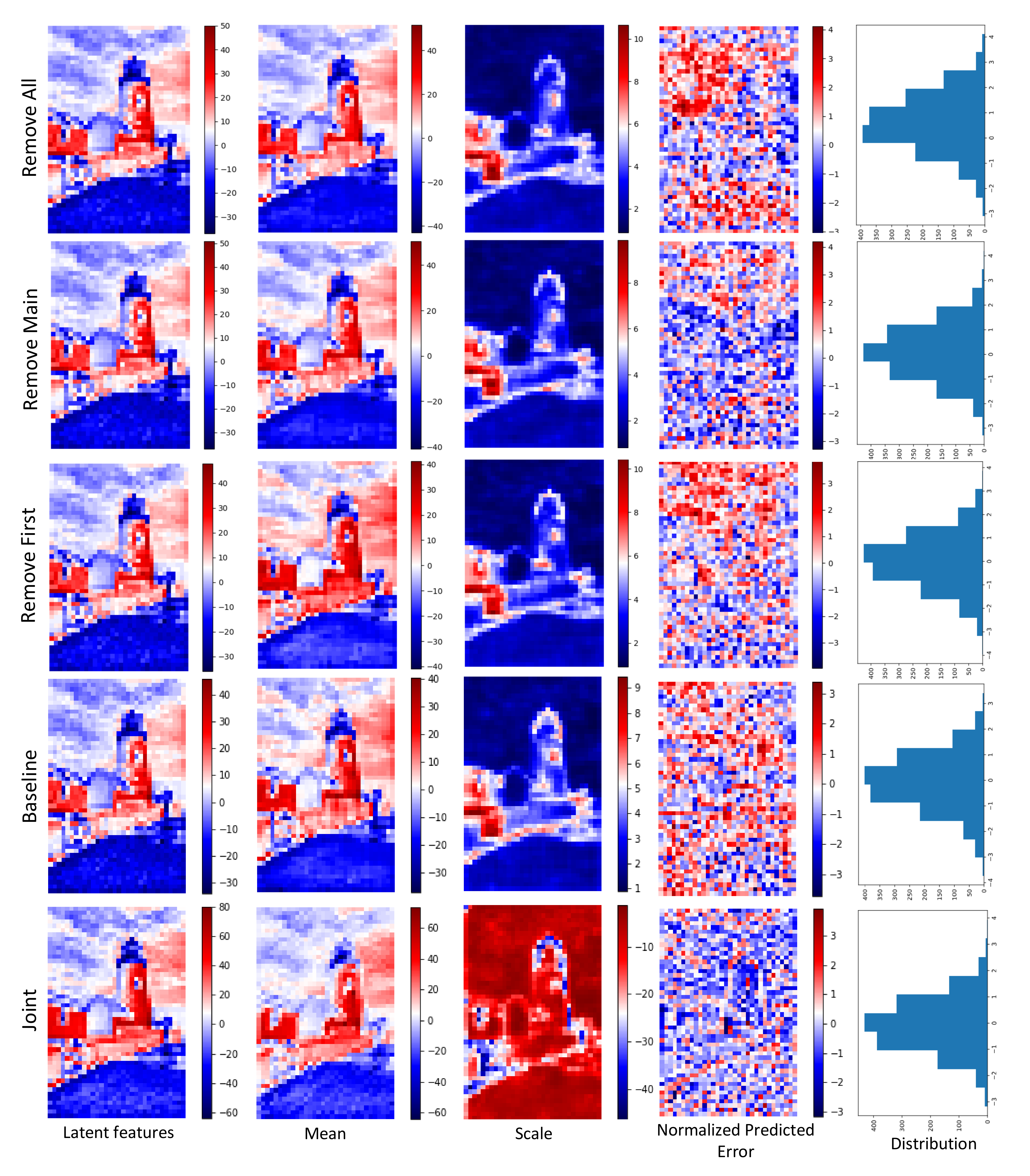}
   \caption{Prediction error with different models at similar bit rate. Column-wisely, it depicts the latent features, the predicted mean, predicted scale,  normalized prediction error ( i.e., $\frac{feature-mean}{scale}$) and the distribution of the normalized prediction error from left to right plots. Each row represents a different model (e.g., various combinations of NLAM components, and contexts prediction). These figures show that with NLAM and joint contexts from hyperprior and autoregressive neighbors, the latent features capture more information (indicated by a larger dynamic range), which leads to larger scales for both predicted mean and scale (standard deviation of features). Meanwhile, the distribution of the final normalized feature prediction error remains compact, which leads to the lowest bit rate. }
   \label{prediction_error}
\end{figure}

 We further compare  conditional context modeling efficiency of the model variants in Fig.~\ref{prediction_error}. As we can see,
with embedded NLAM and joint contexts modeling, our NLAIC could provide more compact latent features,
and less normalized feature prediction error, both contributing to its leading coding efficiency.

In this work, we {\it first} train the ``NLAIC baseline'' models. To train the NLAIC model,
one way is fixing the main and hyperprior encoders and decoders in the baseline model, and updating  only the conditional context model $\mathbb P$.  Compared with the ``NLAIC baseline'', such transfer learning based ``NLAIC'' provides 3\% bit rate reduction at the same distortion. Alternatively, we could use the baseline models as the start point, and refine all the modules in the ``NLAIC'' system. In this way,   ``NLAIC'' offers more than 9\% bit rate reduction over the ``NLAIC baseline'' at the same quality. Thus, we choose the latter one for better performance.

{\bf Impacts of NLAM.} To further delineate the gain due to the newly introduced NLAM, we remove the mask branch in the NLAM pairs gradually, and retrain our framework for performance evaluation. For this study, we use the baseline context modeling (only hyperpriors) in all cases, and use MSE as the loss function and PSNR as the final distortion measurement, shown in Fig.~\ref{model_variants_discuss}. For illustrative understanding, we also provide two anchors, i.e., ``Ball{\'e}2018''~\cite{balle2018variational} and ``NLAIC'' respectively. However, to see the degradation caused by gradually removing the mask branch in NLAMs, one should compare with the NLAIC baseline curve.

 Removing the mask branches of the first NLAM pair in the main encoder-decoders (referred to as ``remove\_first'') yields a PSNR drop of about 0.1dB compared to ``NLAIC baseline'' at the same bit rate. PSNR drop is further enlarged noticeably when removing all NLAM pairs' mask branches in main encoder-decoders (a.k.a., ``remove\_main''). It gives the worst performance when further disabling the NLAM pair's mask branches in hyperprior encoder-decoders, resulting in the traditional variational autoencoder without non-local characteristics explorations (i.e., ``remove\_all'').

\begin{figure}
   \begin{center}
      \includegraphics[scale=0.27]{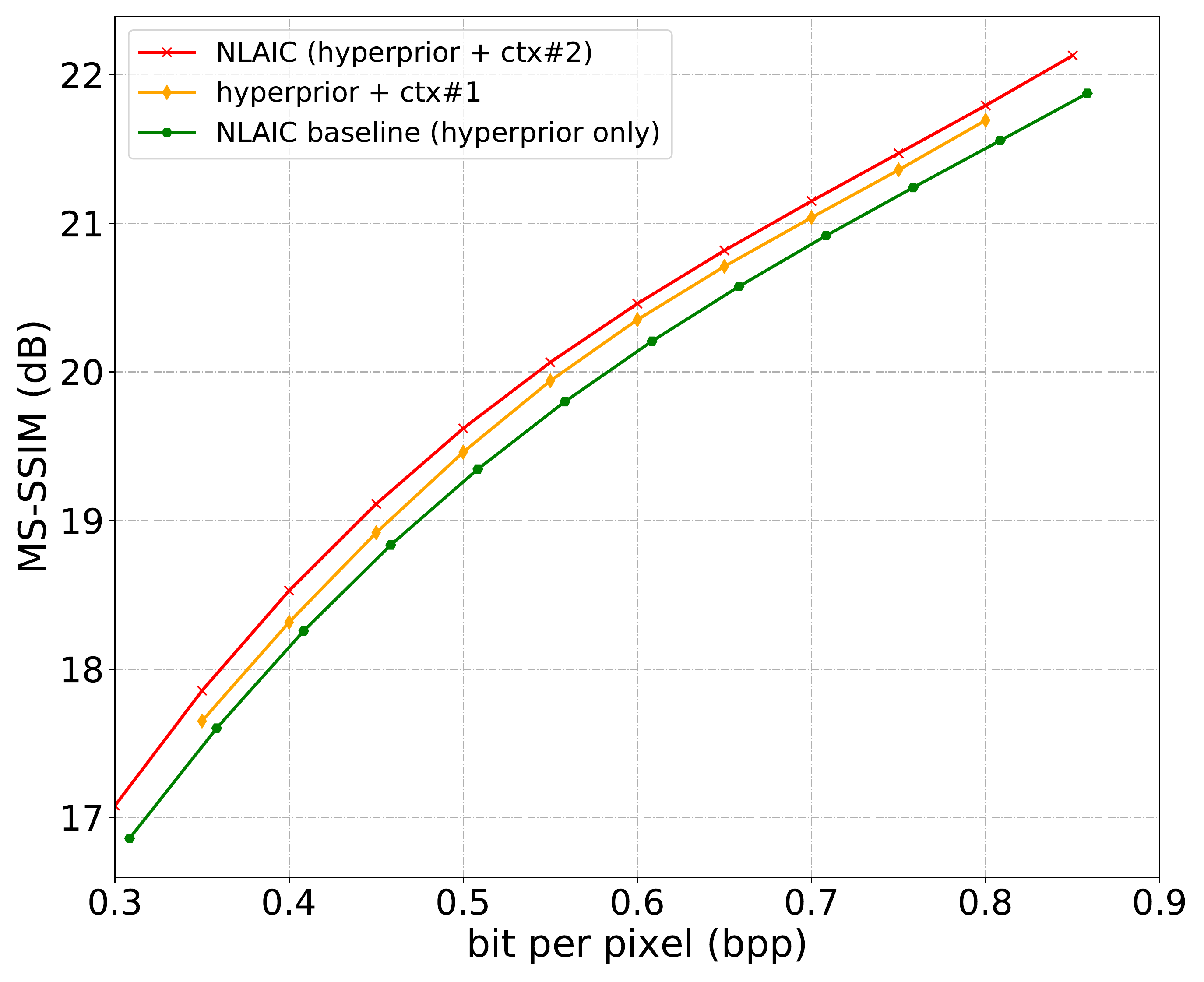}  
   \end{center}
   \caption{{Impacts of Parallel Context Modeling.} Coding Efficiency Illustrations for a variety of conditional probability prediction using hyperprior only (a.k.a., NLAIC baseline), hyperprior + ctx\#1 (a.k.a., contexts w/ channel neighbors only), and hyperprior + ctx\#2 (a.k.a., contexts w/o left neighbor). }
   \label{fig:pixel}
\end{figure}

{\bf Impacts of Parallel Context Modeling.} Parallel 3D CNN-based context modeling is introduced in Section~\ref{ssec:parallel_prob_modeling} to speedup the computational throughput, by removing the prediction dependency on left or left and upper neighbors. 
First, we found that left neighbors have negligible affects to the performance. In this case, we just use contexts without left neighbors in the default NLAIC. 
As shown in Fig.~\ref{fig:pixel}, we have examined three scenarios for conditional probability modeling using, e.g., 1) hyperprior only (a.k.a., NLAIC baseline), 2) hyperprior + ctx\#1 (a.k.a., contexts w/ channel neighbors only), and 3) NLAIC (hyperprior + ctx\#2 (a.k.a., contexts w/o left neighbor)), leading to 6.76\%, 3.73\%, $\approx 0\%$ BD-Rate losses measured by MS-SSIM for respective 1), 2),  3) use cases, when compared with the default NLAIC in with full access to complete information for probability modeling. On the other hand, computational complexity is greatly reduced by enforcing the parallel context modeling.

\begin{figure}[t]
   \centering
   \includegraphics[scale=0.26]{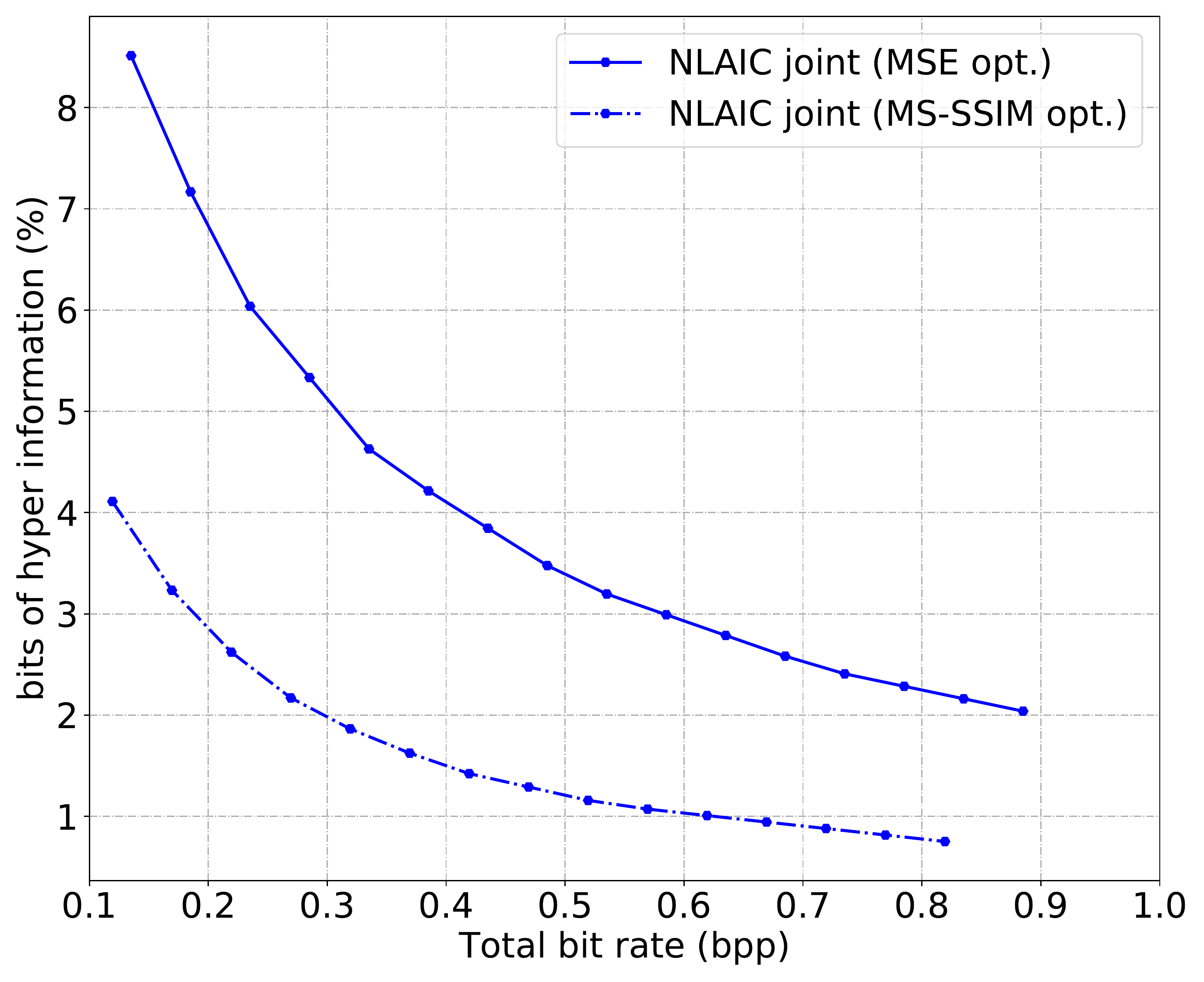}
   \caption{{\bf Percentage of $\hat{z}$.} Bits consumed by $\hat{z}$ in the entire bitstream. 
   The percentage of $\hat{z}$ for MSE loss optimized method is noticeably higher than the scenario using MS-SSIM loss.}
   \label{hyper_bits_comsuming}
\end{figure}

{\textbf{Hyperpriors }$ \bf{\hat{z}}$.} Hyperpriors $\hat{\bf z}$ has noticeable contribution to the overall compression performance ~\cite{minnen2018joint,balle2018variational}. Its percentage decreases as the overall bit rate increases, shown in Fig.~\ref{hyper_bits_comsuming}. The percentage of $\hat{z}$ for MSE loss optimized model is higher than the case using MS-SSIM loss optimization, but still much less than the bits consumed by the latent features.

{\bf Impacts of Unified Model for Variable Rates.} A unified model for variable rates is highly desired for practical application, without resorting to the model re-training for individual rates. As discussed in Sec.~\ref{ssec:unified_model_var_rates}, we have developed quality scaling factors to re-use the models trained at high-bit-rate scenario directly for another set of bitrates.

\begin{figure}
   \begin{center}
      \includegraphics[scale=0.265]{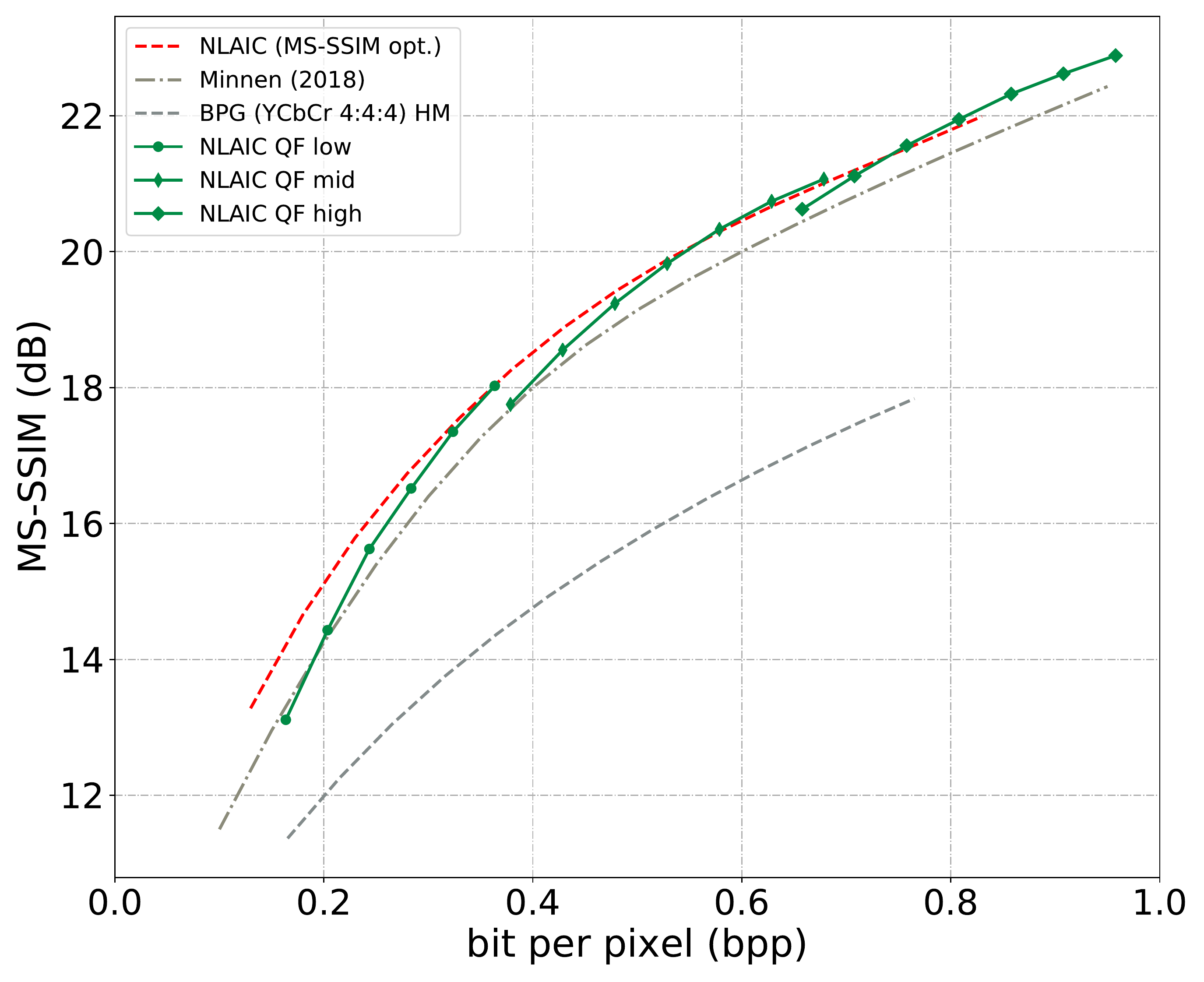}  
   \end{center}
    \caption{RD performance of Multi Bitrates Generation}
    \label{fig:single_model_rd}
   \end{figure}

 Since the original feature maps are shifted, scaled and quantized, it may induce the data distribution variations when having scaling and inverse scaling operations, leading to the degradation of coding efficiency at lower bitrates.
 Usually, coding efficiency degrades larger when the bitrate distance gets further. Thus, we suggest to apply three models instead of a single one to avoid an unexpected performance gap at ultra-low bitrate (assuming the model is trained at the high bitrate), as shown in Fig.~\ref{fig:single_model_rd}, to cover the respective low, medium and high bitrate ranges. For each bitrate range, its model is trained at the highest bitrate. Note that bitrate range may overlap.
 Simulations have revealed that, our approach with three unified models for variable rates, still outperforms the BPG with significant performance margin, and offers the comparable efficiency with Minnen2018~\cite{minnen2018joint} which requires model retraining for each bitrate. Note that the bitrate of hyperpriors are fixed for the entire bitrate range, which causes the performance loss at low bitrates. We can further apply quality factors to the hyper encoder and decoder as well to narrow the gap between our unified model and models that are separately trained.


\section{Concluding Remarks and Future Works} \label{sec:concluding_remarks}

In this paper, we proposed a neural image compression algorithm using non-local attention optimization and improved context modeling (NLAIC). Our NLAIC method  achieves the state-of-the-art performance, for both MS-SSIM and PSNR evaluations at the same bit rate, when compared with existing image compression methods, including well-known BPG, JPEG2000, JPEG as well as the most recent learning based schemes~\cite{minnen2018joint,balle2018variational}.

Key novelties of NLAIC are non-local operations as non-linear transforms to capture both local and global correlations for latent features and hyperpriors, implicit attention mechanisms to adapt more weights for salient image areas, and conditional entropy modeling using joint 3D CNN-based spatial-channel neighbors and hyperpriors. For practical applications, sparse nonlocal operations, parallel 3D CNN-based context modeling, and unified model for variable rates, are introduced to reduce the space, time, and implementation complexity.



For future study, we would like to extend our framework
for end-to-end video compression framework with more priors acquired from spatial and temporal information. Another interesting avenue is to further simplify the models for embedded system, such as fixed-point implementation, platform frienly network structures, etc.

\section{Acknowledgement}
We are very grateful for the constructive comments from anonymous reviewers to improve this manuscript.

\bibliographystyle{IEEEtran}
\bibliography{2019_NVC_p1_v1}

\end{document}